\documentclass[pdflatex,sn-mathphys-num]{sn-jnl}

\usepackage{soul}
\usepackage{graphicx}%
\usepackage{multirow}%
\usepackage{amsmath,amssymb,amsfonts}%
\usepackage{amsthm}%
\usepackage{mathrsfs}%
\usepackage[title]{appendix}%
\usepackage{xcolor}%
\usepackage{textcomp}%
\usepackage[table]{xcolor}
\usepackage{manyfoot}%
\usepackage{booktabs}%
\usepackage{algorithm}%
\usepackage{algorithmicx}%
\usepackage{algpseudocode}%
\usepackage{listings}%
\usepackage{tikz}
\usepackage[normalem]{ulem}
\usepackage[dvipsnames]{xcolor}
\usepackage{graphicx, subcaption}
\graphicspath{{./Figures}}
\usepackage{caption}

\definecolor{DarkBlackGreen}{RGB}{30,80,00}

\theoremstyle{thmstyleone}%
\theoremstyle{thmstyletwo}%

\theoremstyle{thmstylethree}%

\begin{document}
\title[Characterization of Chaotic and Homogeneous coexisting dynamics of a  Memristive Thermo-Controlled MEMS.]{Characterization of Chaotic and Homogeneous coexisting dynamics of a  Memristive Thermo-Controlled MEMS.}


\author*[1,2]{\fnm{N.G.} \sur{Koudafokê}}\email{gilles.koudafoke@unesp.br}

\author[3]{\fnm{Thierry} \sur{Njougouo}}\email{thierry.njougouo@imtlucca.it}

\author[1]{\fnm{Hilda A. } \sur{Cerdeira}}\email{
 hilda.cerdeira@unesp.br}

\author[4]{\fnm{C.H. } \sur{Miwadinou}}\email{clement.miwadinou@imsp-uac.org}


\affil*[1]{\orgdiv{ICTP South American Institute for Fundamental Research, S\~ao Paulo State University (UNESP)}, \orgname{Instituto de F\'{i}sica Te\'{o}rica,Bloco II}, \orgaddress{\street{Rua Dr. Bento Teobaldo Ferraz 271}, \city{Barra Funda}, \postcode{01140-070}, \state{S\~ao Paulo}, \country{Brazil}}}

\affil[2]{\orgdiv{ Institut de Mathématiques et de Sciences Physiques (IMSP)}, \orgname{Université d'Abomey Calavi (UAC)}, \orgaddress{\street{Dangbo}, \city{Porto-Novo},  \state{Ouémé}, \country{Bénin}}}

\affil[3]{\orgdiv{IMT School for Advanced Studies Lucca},  \orgaddress{\street{Piazza San Francesco, 19} \city{}, \postcode{55100}, \state{Lucca}, \country{Italy}}}

\affil[4]{\orgdiv{Ecole Normale Supérieure de Natitingou}, \orgname{Université Nationale des Sciences, Technologies, Ingénierie et Mathématiques d’Abomey},  \country{Bénin}}


\abstract{\begin{abstract}

This work presents the mathematical modelling and numerical investigation of a thermo-controlled Micro-Electro-Mechanical System (MEMS) obtained by coupling an HP memristor with mechanical and electrical resonators. Using the linear drift HP memristor model, the nonlinear electromechanical dynamics are analyzed through Lyapunov exponents, bifurcation diagrams, phase portraits, recurrence plots, Poincar\'e sections, and Fourier spectra. The results reveal parameter-dependent transitions between quasi-periodic and chaotic oscillations together with signatures of coexisting dynamical regimes. A systematic exploration of the intrinsic memristor parameters, namely the ON-state resistance $R_{on}$, the OFF-state resistance $R_{off}$, the oxide thickness $D$, and the ionic mobility $\mu_v$ show that the memristive behavior significantly influences oscillation amplitudes, resonance frequencies, and nonlinear dynamical transitions within the coupled thermo-electro-mechanical system. The state-dependent memristance  dynamically modulates the electromechanical coupling and may redistribute energy between the electrical and mechanical resonators, thereby promoting complex oscillatory responses. In addition, the influence of temperature-sensitive memristive parameters is qualitatively investigated through variations of the ionic mobility and resistive states. The results suggest that moderate thermal variations may affect both the oscillation amplitudes and the dynamical regimes of the system, potentially promoting transitions between quasi-periodic and chaotic behaviors. A comparative discussion with Josephson-junction-based MEMS architectures highlights the operational flexibility and room-temperature compatibility of the HP memristor model for thermo-electro-mechanical applications. These findings suggest promising perspectives for adaptive nonlinear oscillators, thermo-sensitive sensing devices, and chaos-driven electromechanical systems.

\end{abstract}}



\maketitle \footnote{NGK thanks FAPESP---UNESCO-TWAS Project 2024/08346-8 for financial support. HAC
thanks FAPESP grant 2021/14335-0 of the ICTP--SAIFR for partial support.}\\[6pt]

\section{Introduction}
Since its prediction by L. Chua in ``\textit{Memristor --- the missing circuit element.}'' \cite{chua1}, the Memristor has been the subject of extensive scientific investigations, and its memory and nonlinear properties have been demonstrated theoretically, numerically, and experimentally~\cite{Mutlu,strukov,hadis,mazady,hadis2015,hamdi,keshmiri,humood,mokhtar,Belykh,talon}. Owing to their micro- and nanometric dimensions as well as their remarkable electromagnetic properties, memristive devices constitute promising candidates for Micro-Electro-Mechanical Systems (MEMS). MEMS themselves are particularly attractive for high-precision sensing and actuation applications because of their ultra-small dimensions and strong sensitivity to environmental perturbations~\cite{Rouke1}.

Several of our previous studies have employed Josephson junctions in the development of passive and active thermo-sensitive MEMS actuators and sensors~\cite{Koudaf2,Koudaf3}. However, the main limitation of Josephson-junction-based architectures lies in the stringent superconducting conditions required for their operation, especially the need for cryogenic temperatures, which significantly restrict their practical implementation under ambient conditions. Nevertheless, the remarkable nonlinear and dynamical properties of Josephson junctions remain well established and widely documented~\cite{Koudaf2,Koudaf3,Abidi,Ben,Koudaf,Salam1,Salem2,Valkering,Nuznet}. Through their AC and DC Josephson effects, Josephson junctions have been extensively employed in nonlinear and chaotic dynamical systems~\cite{Koudaf2,Koudaf3,Ben,Koudaf,Salam1,Salem2,James}. However, the cryogenic constraints associated with their use may considerably limit the practical scoop of applications of such nonlinear electrodynamic systems.

Motivated by these limitations, the present work investigates the integration of an HP memristor into a thermo-controlled MEMS architecture as an alternative nonlinear memory element operating under standard ambient conditions. The objective is to analyze how memristive nonlinearities influence the coupled thermo-electro-mechanical dynamics of the MEMS and to assess the role of the memristor parameters on the oscillatory regimes of the system. In this context, the HP memristor model was selected because it provides a simple, physically meaningful, and experimentally supported framework for describing nonlinear memory effects while operating under standard ambient-temperature conditions~\cite{Chua2014,Ding2025}. Unlike Josephson junction-based devices, which generally require cryogenic environments to maintain superconductivity, the HP memristor remains fully operational at room temperature, making it suitable for thermo-active MEMS applications~\cite{Yang2013memristive,Pham2022}. Moreover, the HP model captures the essential memristive features, including state-dependent resistance and pinched hysteresis behavior, while keeping the mathematical complexity tractable~\cite{Banchuin2021}. Its compact formulation facilitates coupling with thermo-electro-mechanical equations and allows efficient bifurcation, stability, and Lyapunov analyses~\cite{Ginoux2021,Eftekhari2021}. Furthermore, the HP memristor has become a canonical benchmark in nonlinear dynamics and chaos studies, allowing comparison with previous theoretical and experimental investigations~\cite{Pham2022,Ding2025}.

The electromechanical system investigated in this work consists of an HP memristor coupled to a microbeam resonator and an electrical oscillator $r_0L_0C_0$. In this configuration, the memristor behaves as a central nonlinear element influencing the oscillation amplitudes, resonance frequencies, and dynamical transitions. The oscillatory regimes are characterized through bifurcation diagrams, Lyapunov exponents, recurrence plots, phase portraits, Fourier spectra, and temporal evolution analyses of the principal currents flowing through the MEMS architecture.

The main contribution of this work lies in the development of a thermo-controlled memristive MEMS architecture capable of exhibiting  complex nonlinear dynamics in the electro-mechanical response of the coupled system. By coupling an HP memristor with electrical and mechanical resonators, we show that  the intrinsic memristor parameters $(R_{on}, R_{off}, D,\mu_v)$ may significantly influence oscillation amplitudes, resonance frequencies, and transitions between quasi-periodic and chaotic regimes. The proposed system exhibits thermo-sensitive dynamical behavior through temperature-dependent variations of the memristive parameters, potentially enabling modulation of the electromechanical response. In addition, unlike Josephson-junction-based architectures requiring cryogenic operating conditions, the proposed HP memristor-based MEMS remains compatible with room-temperature operation, thereby offering greater practical flexibility for adaptive sensors, nonlinear oscillators, and chaos-driven MEMS applications.

The remainder of this paper is organized as follows. Section 2 presents the mathematical modelling of the thermo-controlled memristive MEMS and derives the governing electromechanical equations. Section 3 is devoted to the numerical investigation of the nonlinear dynamics, including bifurcation structures, Lyapunov analyses, recurrence plots, phase portraits, and thermo-sensitive dynamical transitions induced by the memristive parameters. Section 4 presents a qualitative comparative analysis of  the dynamical behaviors and thermo-sensitive responses obtained with the proposed  thermo-memristive architecture. Finally, the main conclusions and perspectives of this work are summarized in Section 5.

\section{Mathematical Modelling of MEMS}
\begin{figure}[htp!]
\centering
\includegraphics[width=14cm, height= 6cm,trim=0cm 0cm 0cm 1.5cm,clip]{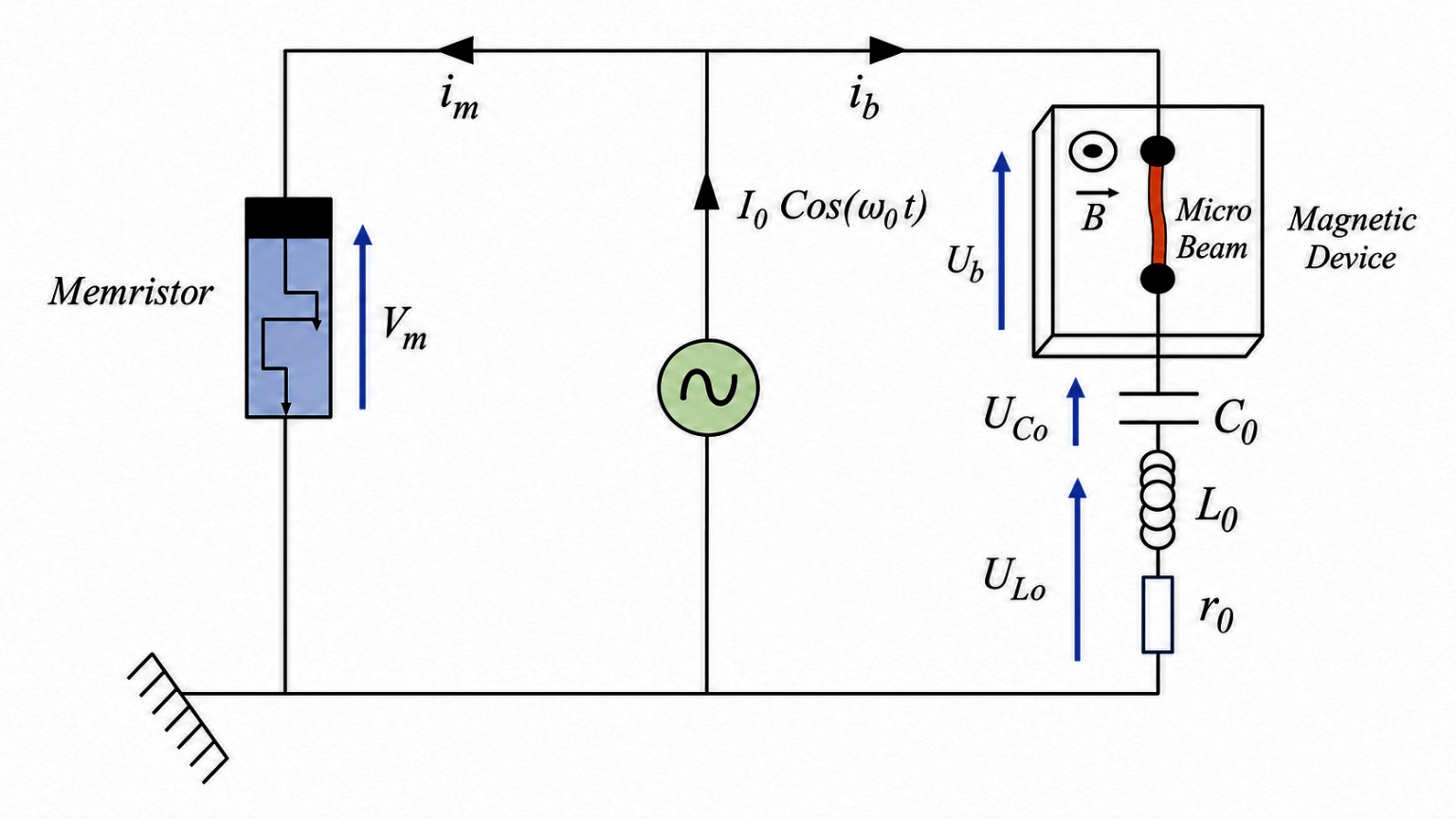}
  
  \caption{Schematic of the memristive MEMS circuit investigated.} \label{Mems1}
\end{figure}

In this section (Subsections \ref{subs1}--\ref{subs3}), we establish the mathematical model used to investigate the MEMS depicted in Fig.~\ref{Mems1}. The circuit consists of an alternating current source with amplitude $I_0$ and excitation frequency $\omega_0$, an HP memristor (Fig.\ref{fig:HP_Mem}), a mechanical resonator (microbeam), and an electrical resonator composed of a resistive inductor $(L_0, r_0)$ and a capacitor $C_0$. The AC source provides energy to the system, while the electrical resonator $(L_0, C_0)$ ensures sustained energy exchange through periodic charging and discharging. The memristor acts as the core control element, regulating the global dynamical response of the MEMS.

The microbeam is used to capture certain physical phenomena, such as magnetic fields and temperature, because of its hypersensitivity to environmental conditions. It may also operate as an ultra-precise micro/nanometric actuator depending on the activation parameters and oscillatory dynamics, while constituting the mechanical element responsible for the activation and transduction of the coupled electromechanical processes within the MEMS architecture. The set of differential equations governing the electrodynamic behavior of the system is derived from Kirchhoff’s laws, the electrical network relations shown in the diagram illustrated in Ref.~\cite{Chaudhary}, and the classical Euler--Bernoulli beam theory, which is introduced in the following subsection.


\subsection{Equations of mechanical oscillations}\label{subs1}
According to \cite{Gerony,Domguia,Kwuimy}, the dynamics of the microbeams can be described by the partial differential equation Eq.~(\ref{eq::mod1}), derived from the classical Euler--Bernoulli beam theory.

\begin{equation}
    EI_y \dfrac{\partial^4 u(z,t)}{\partial z^4}  + \rho A \dfrac{\partial^2 u(z,t)}{\partial t^2} +  \lambda \dfrac{\partial u(z,t)}{\partial t} + NLT = f(t),
    \label{eq::mod1}
\end{equation}
where $E$ denotes the Young’s modulus of the beam material, 
$I_y$ the second moment of inertia of the beam, $\rho$ the volume density, 
$A$ the cross--section of the beam , $f(t)$ the beam activating force, $u(z,t)$ the transverse displacement function, and $\lambda$ the damping coefficient. The  nonlinearity term is assumed to be negligible, i.e., $NLT = 0$.
The actuating force, $f (t)$  is a Lorentz force due to the  current $i_b$ through the beam and is expressed as: 
 \begin{equation}
     f(t) = B \, L \,i_b  =
B \, L \,\left( \dfrac{dq}{dt} + \dfrac{e_b}{r_b}\right)
\label{eq::mod2}
 \end{equation}
with $\vec{B} \, \bot \, L\vec{l}$ 
$
\,\,\,\text{and}\,\,\,
\dfrac{e_b}{r_b}\, =\, \dfrac{- 2\, B\,A}{\mu\,L}\,\int_{o}^{L}\dfrac{\partial u(z,t)}{\partial t}dz.\nonumber\\
$
$L$ is the length of the micro-beam; $\vec{B}$ the magnetic field; Here, $e_b / r_b$ denotes the self--induced current resulting from the beam motion. 
According to Lenz’s law, this induced current acts in opposition to the cause that generate it, which accounts for the negative sign. The term $e_b$ represents the electromotive force  induced in the beam, while the resistance of the beam is expressed as $r_b = \frac{\mu L}{A}$, with$A$ the cross-sectional area, and $\mu$ the resistivity of the micro-beam material. The capacitor $C_0$ has a time-dependent charge $q(t)$, and therefore the total current in the beam  is  the sum of the capacitor charging current $C_0 \, (dq/dt)$ and the self-induced current $e_b / r_b.$ 

By substituting Eq.~(\ref{eq::mod2}) into Eq.~(\ref{eq::mod1}), we obtain the general model expressed in Eq.~(\ref{eq::eqt1}).
\begin{equation}\label{eq::eqt1}
EI_y \frac{\partial^4 u(z,t)}{\partial z^4}  + \rho A \frac{\partial^2 u(z,t)}{\partial t^2} + 
\lambda \frac{\partial u(z,t)}{\partial t} = 
B \, L \,\left[\frac{dq}{dt} -  \frac{2\,B\,A}{\mu\,L}\int_{o}^{L} \frac{\partial u(z,t)}{\partial t} dz.\right] 
\end{equation}
The shape of the modes must satisfy the differential equation and boundary conditions \cite{Gerony} (Chapter 3 Eq.(3.29) ). 
The deflection \, $ u(z, t) $ \, of the beam can then be written as
follows: $ u(z, t) = \sum_{n = 1} ^ {\infty} Z_n (z) \, T_n(t) $ where 
\, $ n $ \, indicates the mode of vibration; \, $ T_n (t) $ \, represents the
generalized coordinate of the amplitudes, and \, $ Z_n (z) $ \, the set of
eigenfunction of the equation:
\begin{equation}
    \frac{\partial^4
	u(z,t)}{\partial z^4} + \frac{\rho A}{E.I_y}\frac{\partial^2
	u(z,t)}{\partial t^2} = 0 .
\end{equation}
In our case, we have $ u(0,t) =u(L,t) = 0$ (boundary conditions) and the set
eigen functions  $Z_n(z)$ are written: 
\begin{equation}
   Z_n(z) = a_n(\cos \xi_n\,z - \cosh \xi_n\,z) +
b_n(\sin\xi_n\,z - \sinh\xi_n\,z) 
\end{equation}
with \,$\xi_n$\,the
solution of the transcendental equation:
\begin{equation}
 \cos \xi_n\,L\,\cosh \xi_n\,L - 1 = 0 .   
\end{equation}

As we focus on the study of the fondamental state, we shall take it for the rest $n=1$.\\
In this study, we will be interested in the dynamics of the middle of the beam ($z =\frac{L}{2}$) in the fundamental mode  \,\,($n=1$). In addition, for this study we will use the normalization made by N. Lobontiu \cite{Lobontiu1,Lobontiu2}. Indeed,
$Z(z) = \eta_1\, Z^*(z^*)$\quad, with \, $z^* = \frac{z}{L}$; \quad $\eta_1 = 16$\quad and \quad $Z^*(z^*) = z^{*^2} (z^* - 1)^2$\\
Using this normalization and the conditions defined above, we obtain:
$Z_1(z) = Z_1(\frac{L}{2}) = 1$
and 
$ \int_{o}^{L}Z_n(z)dz = 0.5333\, L = \dfrac{8}{15} L $

Integrating all these transformations into the equation ~(\ref{eq::eqt1}), we have:
\begin{equation}
	\ddot{T}(t) = -\left(\dfrac{\lambda}{\rho A} + \dfrac{16\,B^2\,L}{15 \mu\,\rho} \right) \dot{T}(t) -  \dfrac{E I_y {\xi_n}^4}{\rho A}T(t)  + \dfrac{B L}{\rho A}\dot{q}(t).
\end{equation}

	\subsection{Equation modelling the variation of  the charge $q(t)$ of capacitor $C_0$}

The voltage $V_m$ at the terminals of a memristor supplied with a current $i_m$ is written as follows \cite{Mutlu}:\\

\begin{equation}
\begin{cases}
V_m(t) = M\!\left[q_m(t)\right]\, i_m(t), \\
M\!\left[q_m(t)\right] = 
\dfrac{d\phi_m(t)}{dq_m(t)} 
= \dfrac{\dot{\phi}_m(t)}{i_m(t)}.
\end{cases}
\end{equation}

Then we can write: (See \cite{Xie,kokate})\\
\begin{equation}
    V_m(t) = \dot{\phi}_m(t) \implies V_m  = R_{on}\left[ \dfrac{R_{off}}{R_{on}}  - \mu_v \dfrac{(R_{off} - R_{on})}{D^2} q_m(t)\right] i_m(t)
\end{equation}

where $\phi_m$ is the flux linkage across the Memristor; 
		$D$ is the Memristor's length;
	$\mu_v$ the Memristor ionic mobility;
		$R_{off}$ the  Memristor resistance in the Off state and
        $R_{on}$ the Memristor resistance in the On state.
        
Applying Kirchhoff's voltage law  (KVL) $V_m - U_{beam} - U_{C_0} - U_{L_0} = 0$, we have:

\begin{equation}
	\ddot{q}(t) = \dfrac{1}{L_0}\dot{\phi}_m(t) - \dfrac{r_p +r_0}{L_0}\dot{q}(t) + \dfrac{16 B L}{15 L_0}\dot{T}(t) - \dfrac{1}{L_0C_0}q(t)
\end{equation}

where $r_{0}$ is the resistance of the Coil ; 
	$C_0$ the self-capacitance of the capacitor,
	$L_0$ the  self-inductance of the coil,
		$I_{0}$ the  amplitude of the excitation current, and  \\
\begin{equation*}\label{}
\dot{\phi}_m(t) = R_{on} \Bigg\{ \frac{R_{off}}{R_{on}} - \mu_v \frac{(R_{of\textbf{}f} - R_{on})}{D^2} 
\Big\{ \frac{I_0}{\omega_0} \sin(\omega_0 t) - q(t) + \frac{16 B A}{15 \mu} T(t) \Big\} \Bigg\}\times
\end{equation*}
\begin{equation}\label{eq4}
\left[ I_0 \cos(\omega_0 t) - \left( \dot{q}(t) - \frac{16 B A}{15 \mu} \dot{T}(t) \right) \right]
\end{equation}

\subsection{Memristor flux equation and governing dynamical system.}\label{subs3}
From the $\phi_m$ equation obtained earlier, and deriving Eq.(\ref{eq4}) once with respect to time, we obtain the following.
\begin{equation}\label{eq5}
\begin{split}
\ddot{\phi}_m(t) = - \mu_v \dfrac{(R_{off} - R_{on}) R_{on}}{D^2}
        \Big\{I_0Cos(\omega_0 t) - \dot{q}(t) + 
        \dfrac{16 B A}{15 \mu}\dot{T}(t)\Big\}^2
        + 	R_{on}\Bigg\{  \dfrac{R_{off}}{R_{on}} -\\ \mu_v \dfrac{(R_{off} - R_{on})}{D^2} 
        \Big\{  \dfrac{I_0}{\omega_0}Sin(\omega_0 t) -
        q(t) + \dfrac{16 B A}{15 \mu}T(t)\Big\} \Bigg\}
        \Bigg\{-I_0\,\omega_0\sin(\omega_0 t)
    - \dfrac{1}{L_0}\dot{\phi}_m(t) 
    + \\
    \Big\{\dfrac{r_p + r_0}{L_0} +
    \dfrac{16\,B^2\,L}{15\,\mu\,\rho}\Big\}\dot{q}(t) - 
    \dfrac{16 B L}{15}\Big\{\dfrac{1}{L_0} + \dfrac{\lambda}{\mu\,\rho\,L} + 
    \dfrac{16\,B^2\,A}{15 \mu^2\,\rho}\Big\}\dot{T}(t) + \dfrac{1}{L_0C_0}q(t)\\
    - \dfrac{16\,B\,E I_y {\xi_n}^4}{15\,\mu\,\rho}T(t)\Bigg\}
\end{split}
\end{equation}
The flow governing the dynamical system is written as follows:\\
\begin{equation}\label{eq8}
	\begin{cases}
			x = \dot{\alpha}_m(\tau)\\
			\dot{x} = - \Delta_{R_m}  \left(I_0\cos(\omega^*_0 \tau) - \omega_1\,Q_0\,z + \sigma_0 \,y\right)^2   +\\
			\sigma_1 \left\lbrace  R^*_0 - \Delta_{R_{on}} \left[ I^*_0 \sin(\omega^*_0 \tau) - Q_0\,\gamma(\tau) +\sigma_2\theta(\tau)\right] \right\rbrace  \times\\
			\left\lbrace -I^*_0\sin(\omega^*_0 \tau)
			- \vartheta_2\,x
			+ \delta_{r_2}\,z
			\right. -\sigma_3\,y 
			+ \sigma_4\gamma(\tau)-
			 \sigma_5\theta(\tau)
			\left.\right\rbrace\\
			y = \dot{\theta}(\tau)\\
			\dot{y}= -\beta_1 y - \beta_2\theta(\tau)  + \beta_3\,z\\
			z=\dot{\gamma}(\tau)\\
			\dot{z} = \vartheta_1\,x - \delta_{r_1}\,z + \eta\,y - \Omega^2_{C_0}\,\gamma(\tau)
	\end{cases}
    \end{equation}

With  $ \tau = \omega_1 t; \, \theta = \dfrac{T(t)}{T_0}; \, \gamma = \dfrac{q(t)}{Q_0}$ \,\, and  \,\, $\alpha_m = \dfrac{\phi_m(t)}{\phi_0} $ where  $\omega_1, T_0, Q_0, \Phi_0$ are scaling coefficients.

$$ \beta_1=\left(\dfrac{\lambda}{\omega_1\rho A } + \dfrac{16 B^2 L}{15\,\omega_1\, \mu\,\rho} \right);\, \, \beta_3 = \dfrac{B\,L\,Q_0}{\rho\,\omega_1\, A\,T_0}; \omega_{C_0}^2=\dfrac{1}{L_0\,C_0};\,\,\vartheta_1 = \dfrac{\phi_0}{L_0\,Q_0\,\omega_1};\,\,
\Omega_{C_0}^2 = \dfrac{1}{ \omega_1^2\,L_0\,C_0};\quad$$ 

$$ \delta_{r_1} = \dfrac{r_p + r_0}{L_0\,\omega_1}; \quad \delta_{r_2} = \left(\delta_{r_0} + \dfrac{16\,B^2\,L}{15\,\mu\,\rho}\right)\omega_1Q_0;\,\, \Delta_{R_{on}} =\mu_v \dfrac{(R_{off} - R_{on})}{D^2};\,\,\,\delta_{r_0} = \dfrac{r_p + r_0}{L_0} $$

$$\Delta_{R_m}=\mu_v \dfrac{(R_{off} - R_{on}) R_{on}}{D^2\,\omega_1^2\,\phi_0};\,\,\, \quad R^*_0 = \dfrac{R_{off}}{R{on}};  \sigma_2 = \dfrac{16 B A T_0}{15 \mu};\,\,\vartheta_0 = \omega_1\phi_0;\,\,\vartheta_2 = \dfrac{ \omega_1\phi_0}{L_0};$$

$$\,\sigma_3 = \dfrac{16 B L\,\omega_1\,T_0}{15}\left(\dfrac{1}{L_0} + \dfrac{\lambda}{\mu\,\rho\,L} +
\dfrac{16\,B^2\,A}{15 \mu^2\,\rho}\right); \beta_2 = \dfrac{E I_y {\xi}^4}{\rho\,\omega_1^2 A};\,\, \sigma_1 = \dfrac{R_{on}}{\omega_1^2\,\phi_0};\,\,\omega^*_0 =\dfrac{\omega_0}{\omega_1}; \,\, I^*_0 = I_0\,\omega_0;\,$$
$$$$
$$\sigma_0= \dfrac{16\,B\,A\,\omega_1\,T_0}{15 \mu}; \,\,\,\, \sigma_4 = \dfrac{Q_0}{L_0C_0}\,;\,\, \sigma_5 = \dfrac{16\,B\,E I_y {\xi}^4\,T_0}{15\,\mu\,\rho}.$$

The main currents of the system are given by:

\begin{equation}
	\begin{cases}
			i_{C_0}(\tau)  = \dot{\gamma} \sim i_b\\
            i_m = \dfrac{\vartheta_0\,x}{  R^*_0 - \Delta_{R_{on}} \left\lbrace I^*_0 Sin(\omega^*_0 \tau) - Q_0\,\gamma(\tau) +\sigma_2\theta(\tau) \right\rbrace }
	\end{cases}
    \end{equation}

	The Jacobian matrix of the system (\ref{eq8}) is written as:
	$$	
	J_x=\begin{pmatrix}
		0&1&0&0&0&0\\
		0&A_x&B_y&C_y&D_z&E_z\\
		0&0&0&1&0&0\\
		0&0&-\beta_2&-\beta_1&0&\beta_3\\
		0&0&0&0&0&1\\
		0&v_1&0&\eta&-\Omega^2_c&-\delta_{r1}
	\end{pmatrix}
	$$
    where\\
		$A_x=-\sigma_1\{R^*_0-\Delta_{R_{on}}{I^*_0\sin(\omega^*_0\tau)-Q_0\gamma(\tau)+
        \sigma_2\theta(\tau)\}\vartheta_2\}}$\\
		\\
		$B_y=-\sigma_1(\Delta_{R_{on}} \sigma_2)\{-I^*_0\sin(\omega^*_0\tau)-\vartheta_2x+\delta_{r2}z-
        \sigma_3y+\sigma_4\gamma(\tau)-\sigma_5\theta(\tau)\}-
		\sigma_5\sigma_1\{R^*_0- \Delta_{R_{on}}[I^*_0sin(\omega^*_0\tau)-Q_0\gamma(\tau)+\sigma_2\theta(\tau)]\}$\\
		\\
		$C_y=-2\Delta_{R_m}\sigma_0\{I_0\cos(\omega^*_0\tau)-\omega_1Q_0z+\sigma_0y\}-
		\sigma_1\sigma_3\{R^*_0-\Delta_{R_{on}}[I^*_0sin(\omega^*_0\tau)-Q_0\gamma(\tau)+\sigma_2\theta(\tau)]\}$\\
		\\
		$D_z=\Delta_{R_{on}} Q_0\sigma_1\{-I^*_0\sin(\omega^*_0\tau)-\vartheta_2x+\delta_{r2}z- \sigma_3y+\sigma_4\gamma(\tau)-\sigma_5\theta(\tau)\}+
	\sigma_4\sigma_1\{R^*_0- \Delta_{R_{on}}[I^*_0\sin(\omega^*_0\tau)-Q_0\gamma(\tau)+\sigma_2\theta(\tau)]\}$\\
		\\
		$E_z=\omega_1Q_0\Delta_{R_m}\{I_0\cos(\omega^*_0\tau)-\omega_1Q_0z+\sigma_0y\}+
		\delta_{r2}\sigma_1\{R^*_0-\Delta_{R_{on}}[I^*_0\sin(\omega^*_0\tau)-
        Q_0\gamma(\tau)+\sigma_2\theta(\tau)]\}$$
    $\\
For the autonomous system ($I_0 =0$), the fixed point is simply the origin $P(x^*,\dot{x}^*,y^*,\dot{y}^*,z^*,\dot{z}^*) = (0,0,0,0,0,0)$.

In the case of sinusoidal excitation, for an applied current $i(t) = I_0 \cos(\omega_0 t)$, the dynamics  become nonlinear and the instantaneous thickness of the conductive region is given by:
\begin{equation}
i_m(t) = \frac{V_m(t)}{M[q_m(t)]} \qquad \text{and } \qquad \frac{dw(t)}{dt} =
\mu_v \frac{R_{{on}}}{D}
\frac{V_m(t)}{{M[q_m(t)]}}
\label{eq15}
\end{equation}
In this case, no simple closed-form analytical solution exists, and the
solution must be obtained numerically. \\

\section{Numerical study and analysis}

\subsection{Physical description  of the HP memristor key parameters employed in numerical analyses}

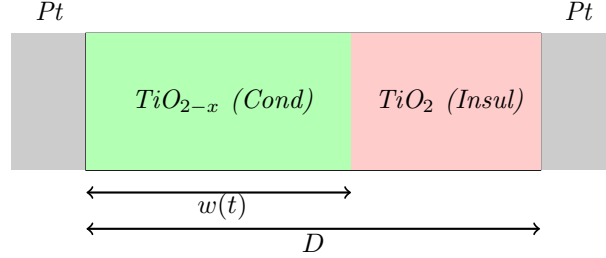
\begin{figure}[h]
    \centering
	\begin{tikzpicture}[x=1cm,y=0.60cm]

	\fill[gray!40] (-4,1.5) rectangle (-3,-1.5);
	\fill[gray!40] (3,1.5) rectangle (4,-1.5);

	\node at (-3.5,2) {\it{Pt}};
	\node at (3.5,2) {\it{Pt}};

	\draw[thick] (-3,-1.5) rectangle (3,1.5);

	\fill[green!30] (-3,-1.5) rectangle (0.5,1.5);
	\node at (-1.2,0) {\it{TiO$_{2-x}$ (Cond)}};

	\fill[red!20] (0.5,-1.5) rectangle (3,1.5);
	\node at (1.8,0) {\it{TiO$_2$ (Insul)}};

	\draw[<->,thick] (-3,-2) -- (0.5,-2);
	\node at (-1.2,-2.3) {\it{$w(t)$}};

	\draw[<->,thick] (-3,-2.8) -- (3,-2.8);
	\node at (0,-3.1) {\it{$D$}};

\end{tikzpicture}

\caption{Physical structure of the HP memristor illustrating the conductive TiO$_{2-x}$ region, the insulating TiO$_2$ layer, and the evolution of the internal state variable $w(t)$ within the active layer thickness $D$.}
\label{fig:HP_Mem}
\end{figure}

The HP memristor model as presented in Fig. \ref{fig:HP_Mem} is characterized by several parameters, three of which are particularly important: the current flowing through the memristor, the voltage across its terminals, and the flux $\phi[q_m(t)]$. These quantities are related through the memristance $M[q_m(t)]$, which  depends on $R_{on}$, $R_{off}$, and $\mu_v$. The parameter $R_{on}$ represents the minimum resistance of the memristor when it is in its conducting state, while $R_{off}$ corresponds to the maximum resistance when it is in its non-conducting state. These parameters play a crucial role in characterizing the behavior of the HP memristor, particularly in applications related to memory devices and neuromorphic computing~\cite{Dmitri,Ji2021,duan,Bao}.
In the HP memristor model \cite{strukov}, the total resistance
results from the series connection of two distinct material regions (See Fig.\ref{fig:HP_Mem}):
\begin{itemize}
	\item a TiO$_{2-x}$ conductive doped region of thickness $w(t)$ as previously defined in Eq.(\ref{eq15})
    \item a insulating non-doped TiO$_2$ region of thickness $D - w(t)$, where $D$ denote the total thickness of the active film.
\end{itemize}

The values of $R_{on}$ and $R_{off}$ are determined exclusively by:
\begin{itemize}
	\item the intrinsic electronic resistivity $\rho_{on}$, \,\, $\rho_{off}$,
	\item the geometry of the device (thickness $D$, surface area $S$),
	\item the density of electron carriers, 
	\item the conduction mechanisms (bands, hopping, and polarons).

\end{itemize}

\begin{equation}
	R_{on} = \rho_{on}\frac{D}{S},
	\qquad
	R_{off} = \rho_{off}\frac{D}{S}.
\end{equation}

Ionic mobility $\mu_v$ describes the drift velocity of  oxygen vacancies 
under an electric field $E_m: v_{ion} = \mu_v E_m.$ Ionic mobility $\mu_v$  is only involved in the dynamical equation~(\ref{eq15}), that is, the time rate of change  of the boundary position $w(t)$,
and not in instantaneous electronic conduction. Thus, in the ideal linear HP model,
$R_{on}$ and $R_{off}$ do not depend on $\mu_v$,
as they are related to electron transport,
while $\mu_v$ governs only the kinetics of ion migration.

\begin{figure}[htbp]
    \centering
    \begin{subfigure}[b]{0.45\textwidth}
        \centering
        \includegraphics[width=6cm, height=4.5cm,trim=0cm 0cm 0cm 2.5cm,clip]{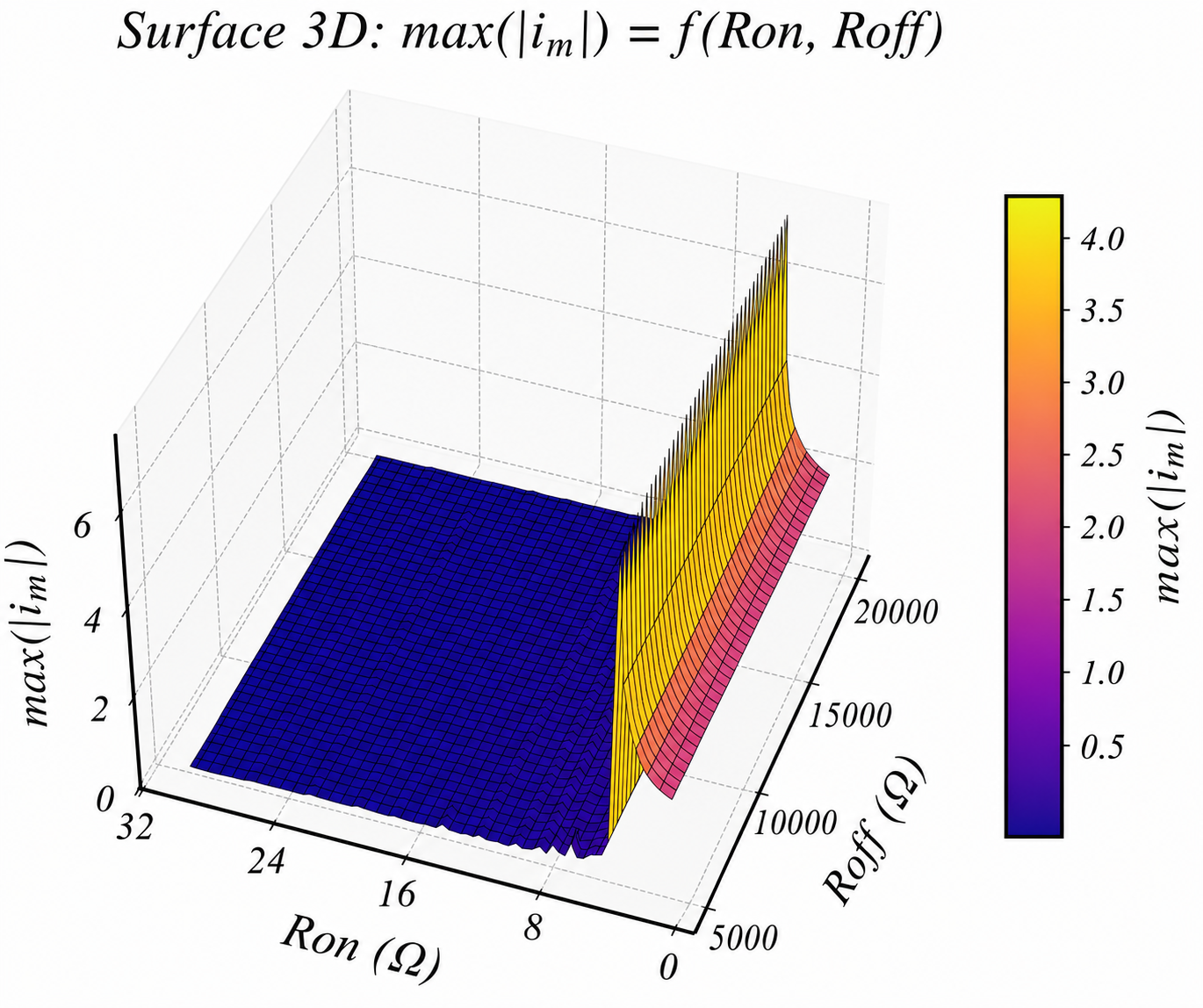}
        \caption{Variation of\,\,$i_m = f(R_{on}, R_{off})$}
    \end{subfigure}
    \hfill
    \hspace{-2cm}
    \begin{subfigure}[b]{0.45\textwidth}
        \centering
        \includegraphics[width=6cm, height=4.5cm,trim=0cm 0cm 0cm 2.5cm,clip]{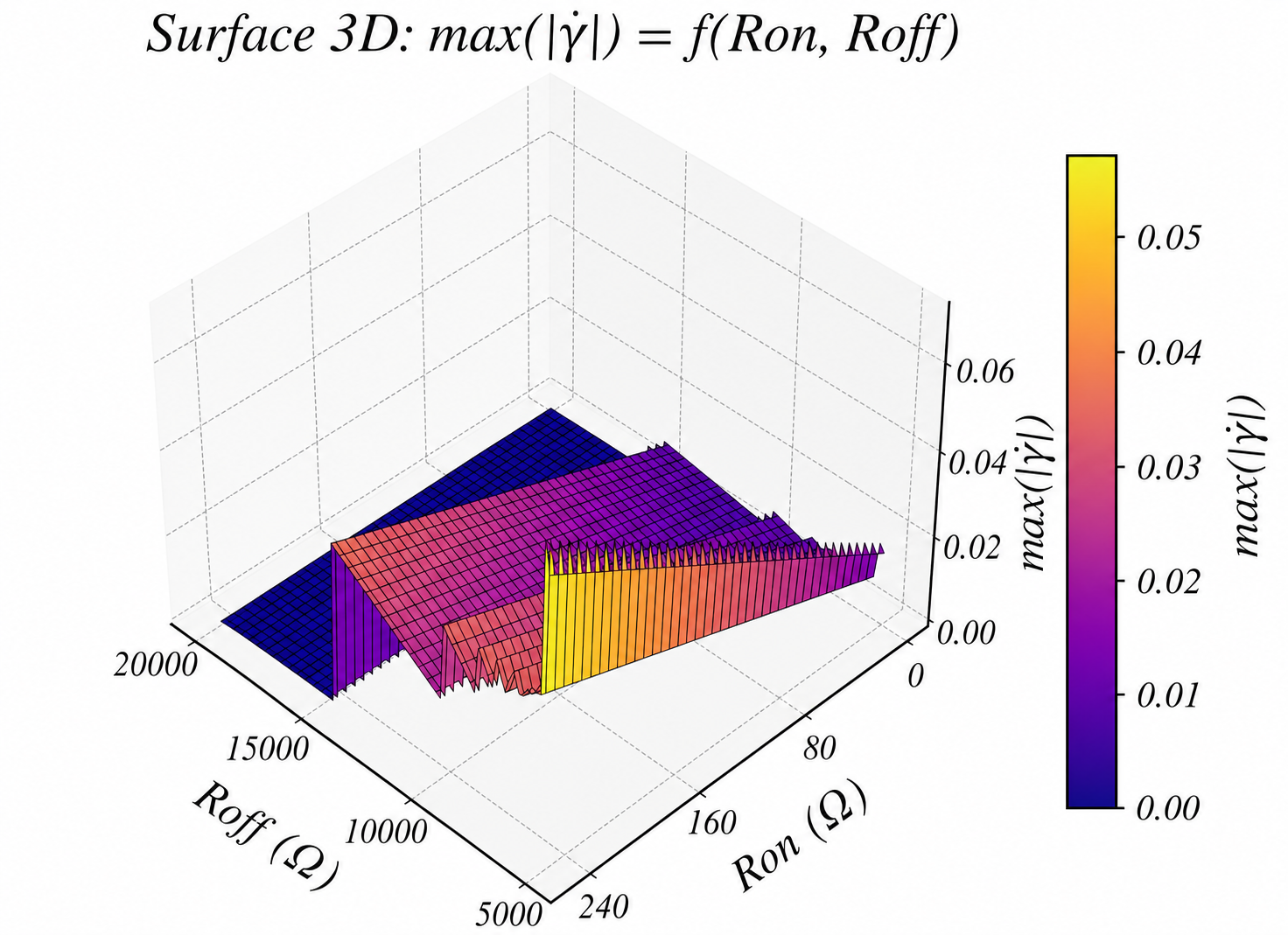}
        \caption{Variation of\,\,$i_{C_0} = \dot\gamma=f(R_{on}, R_{off})$}
    \end{subfigure}
    \caption{System currents as a function of $R_{on}$ and $R_{off}$ : (a) The memristor current  $i_m$  and (b) The current $i_b$ through the beam and capacitor $C_0$. Fig. \ref{fig4}(b) is rotated with respect to Fig.\ref{fig4}(a), to better visualize the current distribution. $\, D = 5nm;\,\, \mu_v= 10^{-10} cm^2 s^{-1}V^{-1}, I_0=2 mA$.}\label{fig4}
\end{figure}

	\qquad
	\qquad
Accordingly, Figs.~\ref{fig4}--\ref{fig15} were generated to investigate the various electromechanical oscillatory regimes exhibited by the proposed memristive MEMS within parameter domains identified from the Lyapunov exponent diagrams. In most of these analyses, the ionic mobility $\mu_v$ is assumed constant (without intrinsic or extrinsic thermal activation) in order to isolate the influence of the intrinsic memristor parameters $R_{on}$ and $R_{off}$ on the nonlinear electrodynamic behavior of the system. Variations of these parameters significantly modify the internal resistive state of the memristor, given by:
\begin{equation}
R(t)=R_{on}w(t)+R_{off}(1-w(t)),
\qquad \text{with} \qquad
R_{on}\leq R(t)\leq R_{off},
\label{eqR}
\end{equation}
together with the accumulated  memristive charge $q_m(t)$, thereby influencing the nonlinear energy transfer mechanisms within the coupled thermo-electro-mechanical MEMS (Figs.~\ref{fig4}--\ref{fig8}, \ref{fig11.1}--\ref{fig15}). As a result, the system may undergo transitions between distinct oscillatory regimes ranging from quasi-periodic to chaotic dynamics. Nevertheless, a dedicated Lyapunov analysis involving variations of $\mu_v$ was also performed to evaluate the influence of ionic migration kinetics on the emergence and modulation of chaotic dynamics(Fig.~\ref{fig11}). In this framework, the memristor does not merely behave as a passive memory element, but could rather as an active nonlinear regulator capable of dynamically redistributing energy between the electrical and mechanical resonators. Through the modulation of the memristive parameters, the device directly influences resonance amplification, oscillatory stability, and the emergence of nonlinear dynamical states within the MEMS architecture (Figs.~\ref{fig4}--\ref{fig15}).
It should be emphasized that the ionic mobility $\mu_v$ in oxide-based memristive devices strongly depends on several physical factors, including oxygen vacancy concentration, initial doping conditions, defect density, fabrication process, local electric field, and temperature~\cite{Khan2024,garcia2016spice,singh2018temperature}. Consequently, the effective mobility may vary over several orders of magnitude depending on the material composition and operating conditions. In the present work, $\mu_v$ was varied within the interval $10^{-12}$ to $5\times10^{-10}\,\mathrm{m^2V^{-1}s^{-1}}$, consistent with the order of magnitude considered in the original HP memristor model proposed by Strukov \textit{et al.}~\cite{strukov}. Furthermore, slightly extended values of $\mu_v$ were considered (Figs.~\ref{fig4D} and~\ref{fig92}) in order to explore a broader thermo-electro-mechanical nonlinear dynamical domain and to evaluate the sensitivity of the oscillatory regimes to enhanced ionic transport effects. Such a parametric approach remains physically plausible owing to the strong variability of ionic mobility reported in oxide-based memristive materials. The present work mainly highlights the indirect influence of temperature on energy redistribution and nonlinear dynamical mode reconfiguration. It should be noted that, although the present study does not explicitly establish the thermal operating limits of the proposed MEMS architecture, recent investigations on TiO$_2$-based memristive devices generally report operating temperature ranges extending from approximately $200$ to $450$\,K~\cite{singh2018temperature,Khan2024,Vaidya2021a,Vaidya2021b}.

The numerical study begins with an analysis of the distribution of the principal currents of the system as a function of the HP memristor parameters $R_{on}$ and $R_{off}$. Subsequently, a comprehensive nonlinear dynamics investigation is carried out using bifurcation diagrams, Lyapunov exponents, Poincar\'e sections, recurrence plots, Fourier amplitude spectra, phase portraits, and time evolution responses. This analysis reveals the emergence of chaotic and quasi-periodic regimes induced by variations in the memristor parameters. Furthermore, a control strategy is proposed to regulate these dynamical behaviors. Finally, a brief examination is conducted to assess the plausible electromechanical responses of the system under moderate thermal variations affecting the HP memristor.

Since the system proposed in this work is purely excited by a sinusoidal current and to assess the influence of memristor on the system's oscillation modes,  it is therefore relevant  to map the behavior of the system's main currents as a function of $R_{off}$ and $R_{on}$. The behavior of the memristor current $i_m$ was numerically evaluated in Fig.~\ref{fig4} by simultaneously varying $R_{on}$ and $R_{off}$. This allowed us not only to confirm  the expected nonlinear nature of the device but also to verify the consistency  of the mathematical model through the conformity of the current dynamics with Kirchhoff's laws. The numerical analysis is performed in Python using Runge-Kutta  RK45 (Dormand–Prince), as implemented in $scipy.integrate.solve\_ivp$. All axes representing dimensionless  variables are shown without units during the numerical analysis. For brevity, certain parameter symbols whose unit are lenghty, such as $\mu_v$, are defined in Table \ref{tab:table1r} and recalled in the figure captions. The various curves shown in this section are obtained with the following parameter values: $L = 200\, \mu m; \,\, l = h = 1\,\mu m;\,\, \mu = 0.2582\,\, \Omega.m;\,\, E = 150.10^{9} \,Pa;\,\,\rho = 2330 \,kg.m^{-3}; \, D = 5\,nm;\,\, \mu_v= 10^{-10}\, (cm^2 s^{-1}V^{-1}.);\,\, L_0 = 50\, mH;\,\,r_0 = 5\, \Omega;\,\,C_0 = 50 \,pF;\,\,\omega_0=\omega_b/35;\,\,\omega_1=\omega_b = 2\pi\times1.05\times \sqrt(\dfrac{E}{\rho})\dfrac{h}{L^2}$ with $\omega_b$ the natural frequency of the beam (see \cite{strukov,Ji2021, Mohanty, Etaki, Ekinci1, Ekinci2}). The intensity of the magnetic field $\vec{B}$ is $B=1\, T$ (see \cite{Ekinci1,Ekinci2}). The initial conditions are chosen as $\alpha_0 = 0.2; x_0 = 0.05; \theta_0 =0.05; y_0 =0.35;\gamma_0 = -1; z_0 = 0.25$.

\begin{table}[h]
\caption{Nature and units of the physical parameters of the system used in the numerical study.}\label{tab:table1r}%
\begin{tabular}{@{}llll@{}}
\toprule
\textbf{Designation} & \textbf{Nature} \& \textbf{Unit} & \textbf{Designation} & \textbf{Nature} \& \textbf{Unit}\\
\midrule
\textbf{$R_{on}$} & Scalar  parameter ($\Omega$) & \textbf{$(\alpha_m;\,\, \dot{\alpha}_m)$} & Dimensionless memristor variables\\

\textbf{$R_{off}$} & Scalar  parameter ($\Omega$) & \textbf{$(\theta;\,\, \dot{\theta})$} &Dimensionless beam variables\\

\textbf{$\mu_{v}$} & Scalar  parameter ($cm^2 s^{-1}V^{-1}$) & \textbf{$(\gamma;\,\, \dot{\gamma})$} & Dimensionless $C_0$ charge variables\\

\textbf{$D$} & Scalar  parameter ($nm$) & \textbf{$i_{m}$} & Dimensionless memristor current \\

\textbf{$I_{0}$} & Parameter with algebraic value ($mA$) & \textbf{$\tau$} & Dimensionless time\\

\textbf{$\mu$} & Scalar  parameter ($\Omega.m$) & \textbf{$i_{C_0} = \dot{\gamma} \sim i_b$} & Dimensionless capacitor $C_0$ current\\

\botrule
\end{tabular}
\end{table}

A brief analysis of the first diagram in Fig.~\ref{fig4} highlights the nonlinear properties of the memristor. For very low values of $R_{on}$, a maximum current flows through the memristor (Fig.~\ref{fig4}(a)), while a minimum current $i_b$ flows through the capacitor $C_0$ (Fig.~\ref{fig4}(b)). When $R_{on}$ is very small, current saturation occurs in the memristor. As $R_{on}$ increases, the current $i_b$ flowing through the beam and charging the capacitor $C_0$ also increases. The approximately  nonlinear variation of $\max(\dot{\gamma})$ with the ratio $\dfrac{R_{off}}{R_{on}}$ suggests a variable dynamics of the current $i_b$ flowing through the beam, leading to increased mechanical amplitudes. This behavior occurs as the memristor current $i_m$ reaches a peak for very small values of $R_{on}$ before reaching an apparent saturation plateau  as $R_{on}$ increases (see Fig.~\ref{fig4} and Fig.~\ref{fig6}(a) and \ref{fig6}(b)). It should be noted that for values of $R_{off}$  between $10\,\mathrm{k}\Omega$ and $15\,\mathrm{k}\Omega$, which correspond to the preferred operating range of the HP memristor for periodic oscillations (see~\cite{radwan}), the current $i_b$ flowing through the beam exhibits an almost linear dependence on the ratio $\dfrac{R_{off}}{R_{on}}$. These two diagrams confirm that the mathematical model is consistent with the physical laws and properties of the memristor used.

In order to evaluate the general dynamics of the system and the actual effects of the other memristor parameters on the electromechanical dynamics, we use $R_{off}$ in the largest range of linearity of $\dot\gamma$ $(10 k\Omega \leq R_{off} \leq 15 k\Omega$, see Fig.~\ref{fig4} (b)), corresponding well to the range of values of $R_{off}$ for the regular oscillations of the HP memristor (see \cite{radwan}). This allows the dynamics to remain consistent throughout this range of $R_{off}$.

 \subsection{Bifurcation analysis and nonlinear dynamics of the memristive MEMS without thermal effects}
Given the strong sensitivity of the system currents to variations in $R_{on}$ (Fig.~\ref{fig4}), the bifurcation structure of the memristor current $i_m$, together with the corresponding maximum Lyapunov exponent (LE) distributions, was investigated as functions of $R_{on}$ and the excitation current amplitude $I_0$. Figure~\ref{fig42} reveals two main dynamical regions. The first corresponds to a central domain characterized by alternating chaotic and quasi-periodic oscillations, whose width progressively decreases as $R_{on}$ increases. The second region, approximately symmetric with respect to the central axis, is predominantly associated with quasi-periodic electromechanical oscillations. In addition, for $\lvert I_0 \rvert \geq 2\,\mathrm{mA}$, the oscillatory dynamics tend to become more regular. The maximum Lyapunov exponent remains close to zero in these regions, suggesting the predominance of quasi-periodic rather than purely periodic oscillatory regimes. The progressive reduction of the chaotic region with increasing $R_{on}$ suggests that larger ON-state resistances tend to attenuate the nonlinear electromechanical interactions responsible for complex oscillatory responses. This behavior may be related to a reduction in the effective energy transfer between the electrical and mechanical subsystems, thereby promoting more regular dynamical responses. Furthermore, the approximate symmetry of the bifurcation structures with respect to $I_0=0$ indicates that the system exhibits comparable dynamical responses for opposite excitation polarities within the investigated parameter range.

\subsubsection{ Chaotic and homogeneous behavior in memristor-based MEMS}
\subsubsection*{\textbf{3.2.1.1\,\,Alternating electromechanical dynamical regimes in the positive Lyapunov exponent region}}
\begin{figure*}[!htb]
    \centering 
    \begin{subfigure}[b]{0.8\textwidth}
        \centering
        \includegraphics[width=12cm, height=5.5cm]{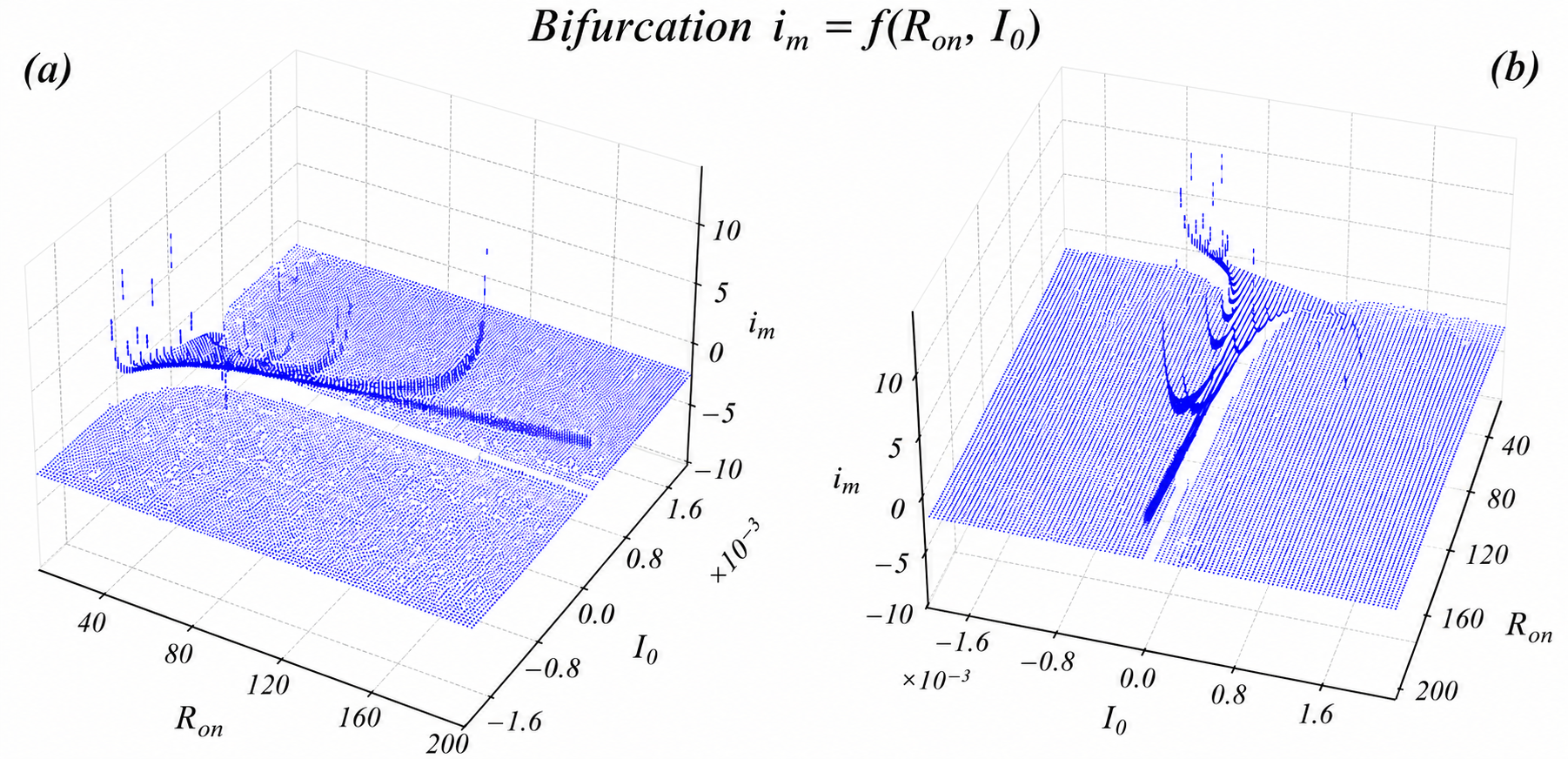}
        \label{fig:graphe1}
    \end{subfigure}
    \vspace{0.1cm}
    \begin{subfigure}[b]{0.8\textwidth}
        \centering
        \includegraphics[width=6cm, height=4.5cm]{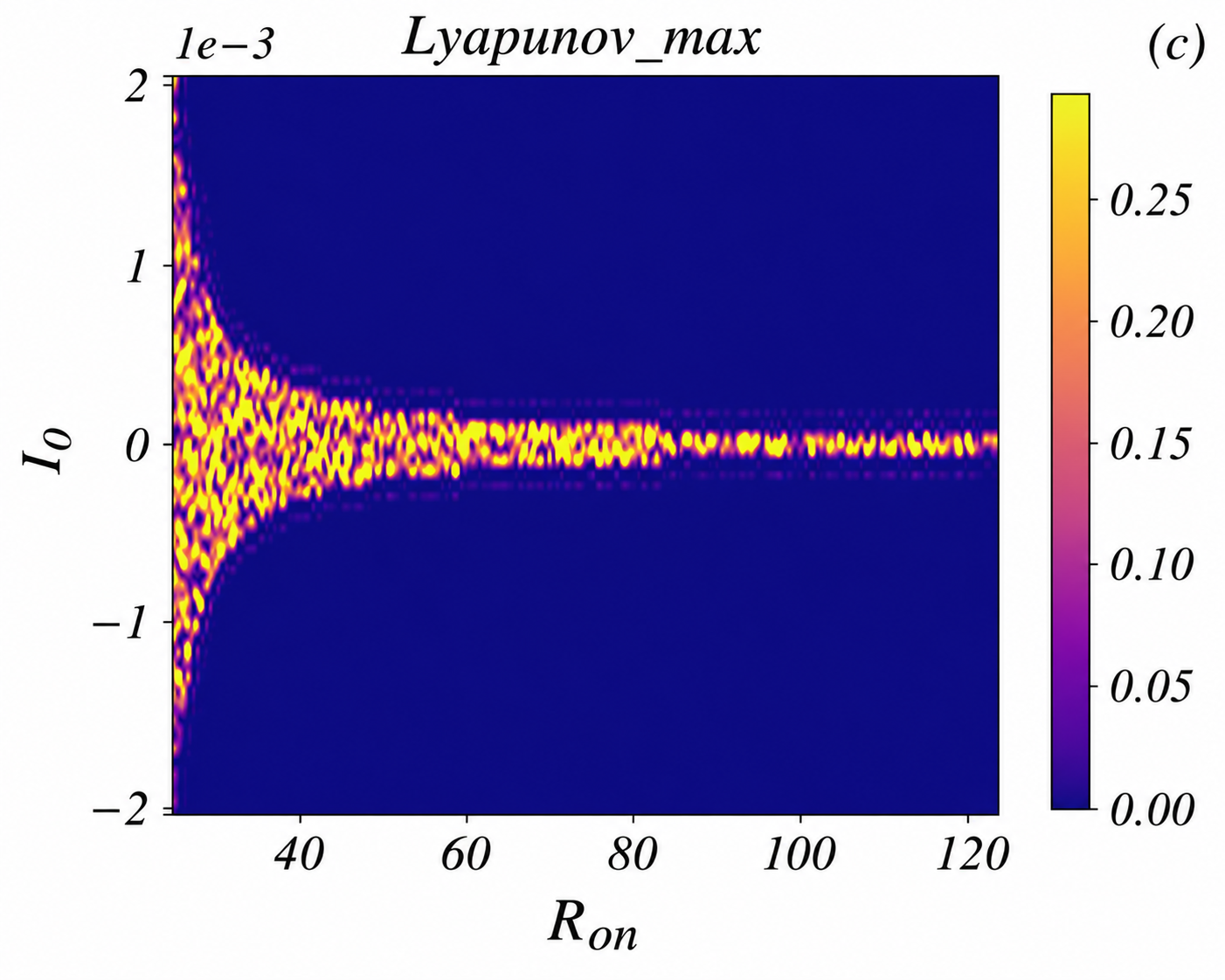}
        \label{fig:graphe2}
    \end{subfigure}
    \caption{Bifurcation diagrams and maximum Lyapunov exponent map of the memristive MEMS as functions of $R_{on}$ and $I_0$ for $\mu_v =10^{-10}\,{cm^2\,s^{-1}\,V^{-1}}$, $R_{off} =12\,{k}\Omega$, and $D=5\,{nm}$. Panels (a) and (b) correspond to two viewing orientations of the bifurcation structure of the memristor current $i_m$, whereas panel (c) shows the associated Lyapunov map identifying regular and chaotic dynamical regions.}
    \label{fig42}
\end{figure*}

\begin{center}
\begin{figure}[htt]
        \hspace{0cm}\includegraphics[width=12cm, height=4.5cm,trim=0cm 0cm 0cm 1.8cm,clip]{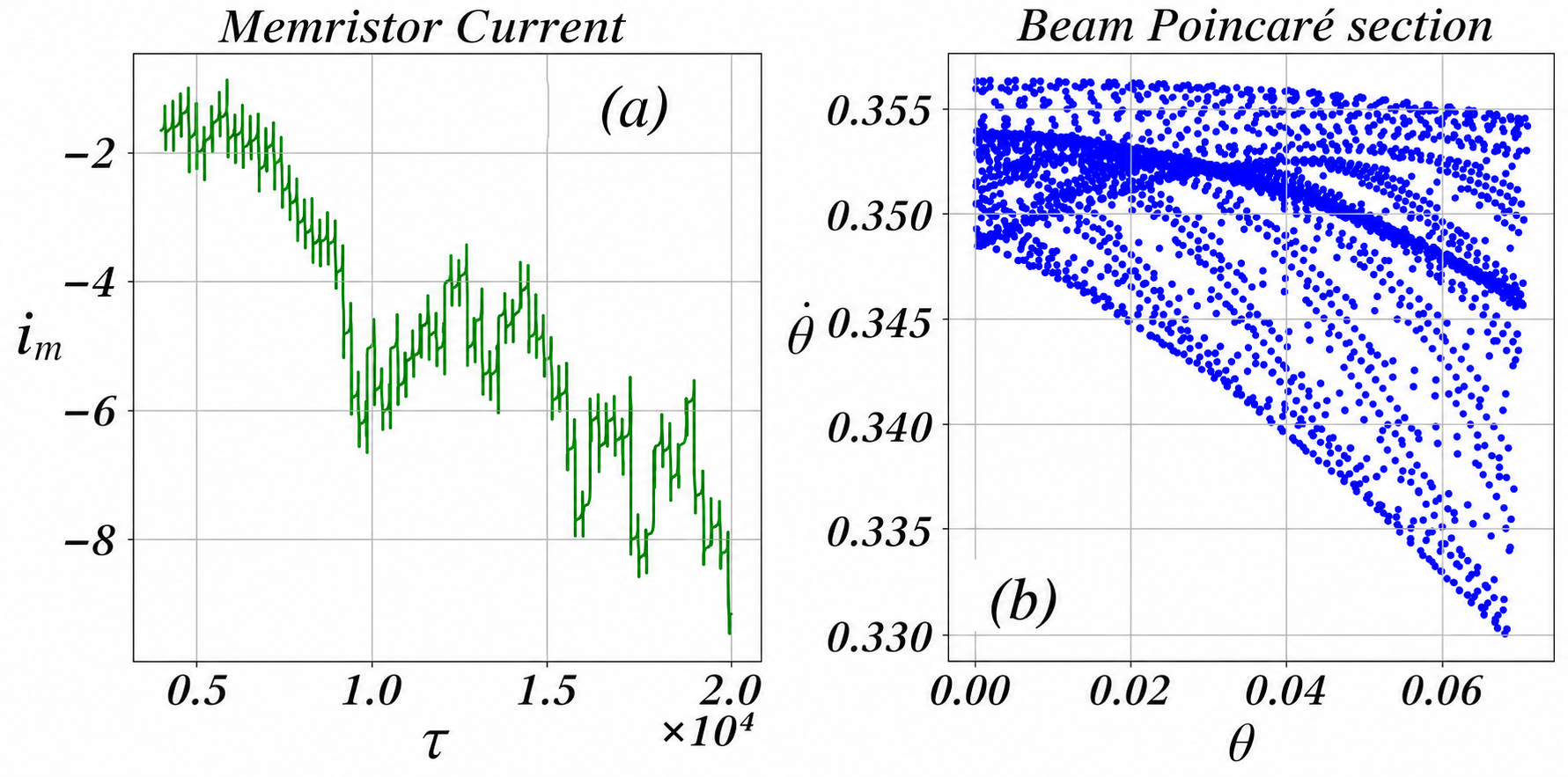}
    \caption{Temporal memristive response and associated beam Poincar\'e section highlighting alternating nonlinear electromechanical dynamics.$(a):$ Memristor current time evolution. $(b)$: Beam Poincaré section for \quad$D= 5 nm, \,\mu_v = 10^{-10} cm^2 s^{-1}V^{-1}$\,\,, $I_0 = -0.5 mA$\,\,,  $R_{on} = 20 \Omega$, \,\,and\,\, $R_{off} =12\,{k}\Omega$.} \label{fig91}   
  \end{figure}
\end{center}

The dynamical responses illustrated in Fig.~\ref{fig91} were obtained by selecting a parameter point located within the region of positive Lyapunov exponents. The time series were computed over $20\,000$ time units $(\tau)$, after discarding the initial $5\,000$ time units to eliminate the transient effects.
The memristor current exhibits irregular and aperiodic temporal oscillations, indicative of a sensitive dependence on initial conditions. Furthermore, the Poincaré section of the beam oscillations reveals a highly scattered and unstructured point distribution, characteristic of a chaotic attractor.
These observations confirm the presence of chaotic electromechanical oscillations, in agreement with the bifurcation and Lyapunov analyses reported in Fig.~\ref{fig42}. 

The temporal coexisting dynamical behavior is supported by both the temporal evolution of the memristor current and  the beam Poincar\'e section. In Fig.~\ref{fig91}(a), the memristor current exhibits successive intervals of relatively organized (homogeneous) oscillations separated by irregular amplitude modulations, suggesting intermittent transitions between distinct oscillatory responses. This behavior is further confirmed by the Poincar\'e map in Fig.~\ref{fig91}(b), where locally concentrated trajectories coexist with strongly dispersed point distributions. The presence of these alternating organized and scattered regions indicates intermittent switching between quasi-periodic and more complex chaotic electromechanical dynamics. Such alternation may be associated with the temporal evolution of the effective memristive resistance
$R(t)$ (Eq.~\ref{eqR}),
whose dynamics continuously modulate the nonlinear electromechanical coupling.

 \subsubsection*{\textbf{3.2.1.2\,\,Quasi-periodic dynamics in the uniform LE domain}}

\begin{center}
\textbf{{\it{Numerical analysis of the dynamics of the HP memristor current $i_m$}}}\\
\end{center}

\begin{figure}[h]
\centering

\begin{subfigure}{0.48\linewidth}
    \centering
    \includegraphics[width=5cm, height=3cm,trim=0cm 0cm 0cm 3cm,clip]{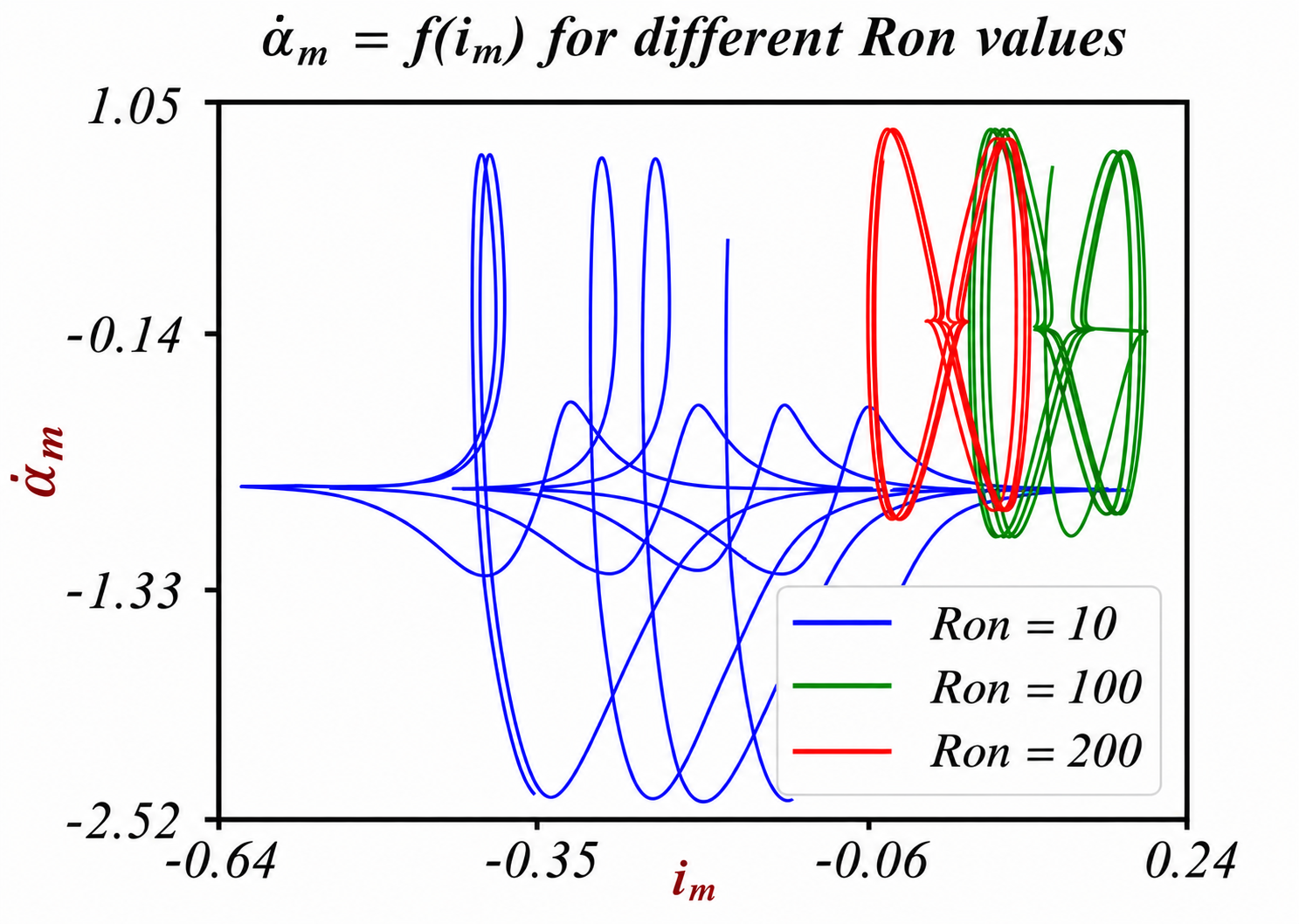}
    \caption{\textit{Voltage-Current Characteristics $\dot{\alpha}_m (i_m)$}}
\end{subfigure}
\hfill
\begin{subfigure}{0.48\linewidth}
    \centering
    \includegraphics[width=5cm, height=3cm,trim=0cm 0cm 0cm 3cm,clip]{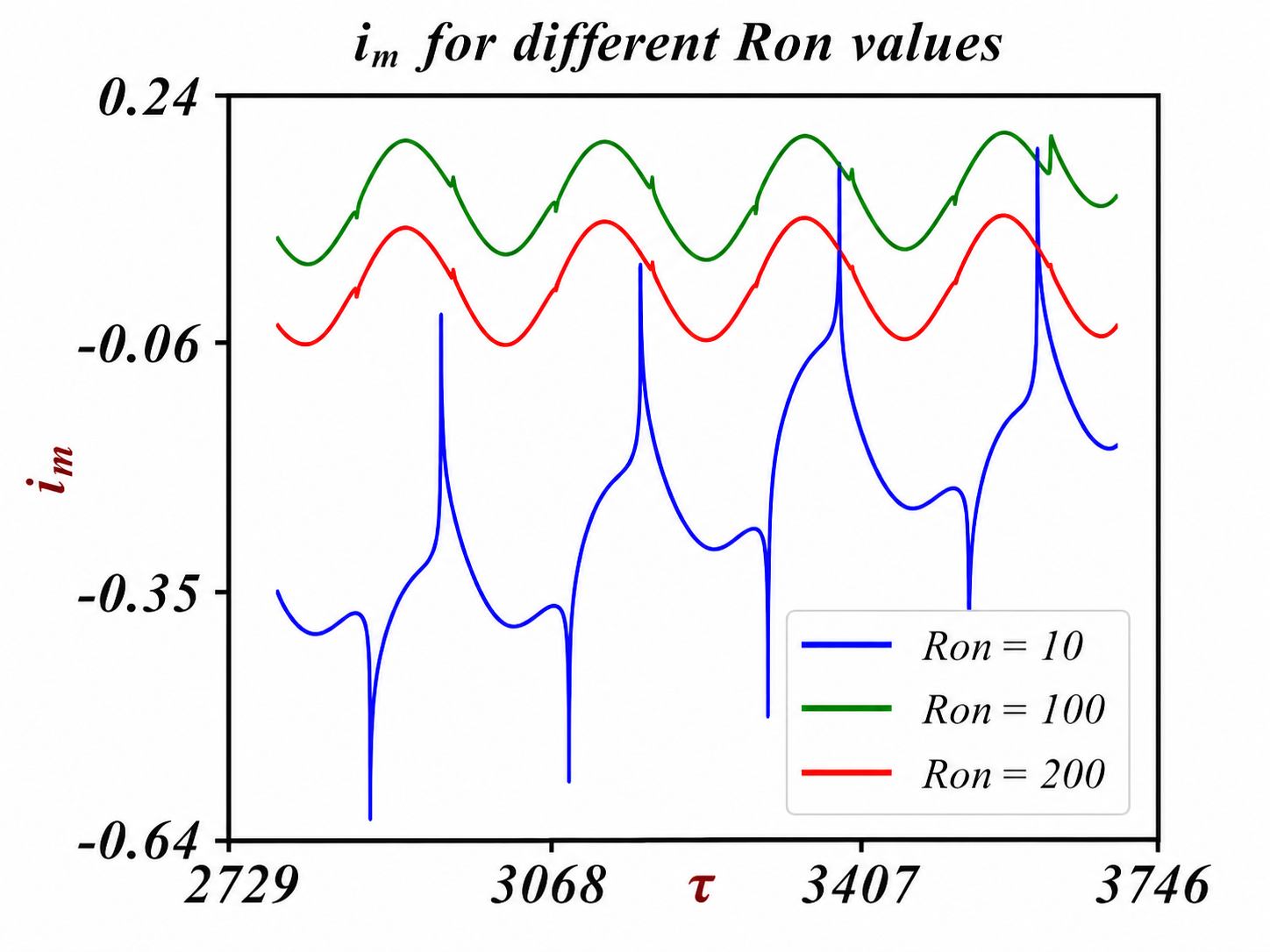}
    \caption{ Memristor current $i_m$ time evolution}
\end{subfigure}
\begin{subfigure}{0.48\linewidth}
    \centering
    \includegraphics[width=5cm, height=3cm,trim=0cm 0cm 0cm 3cm,clip]{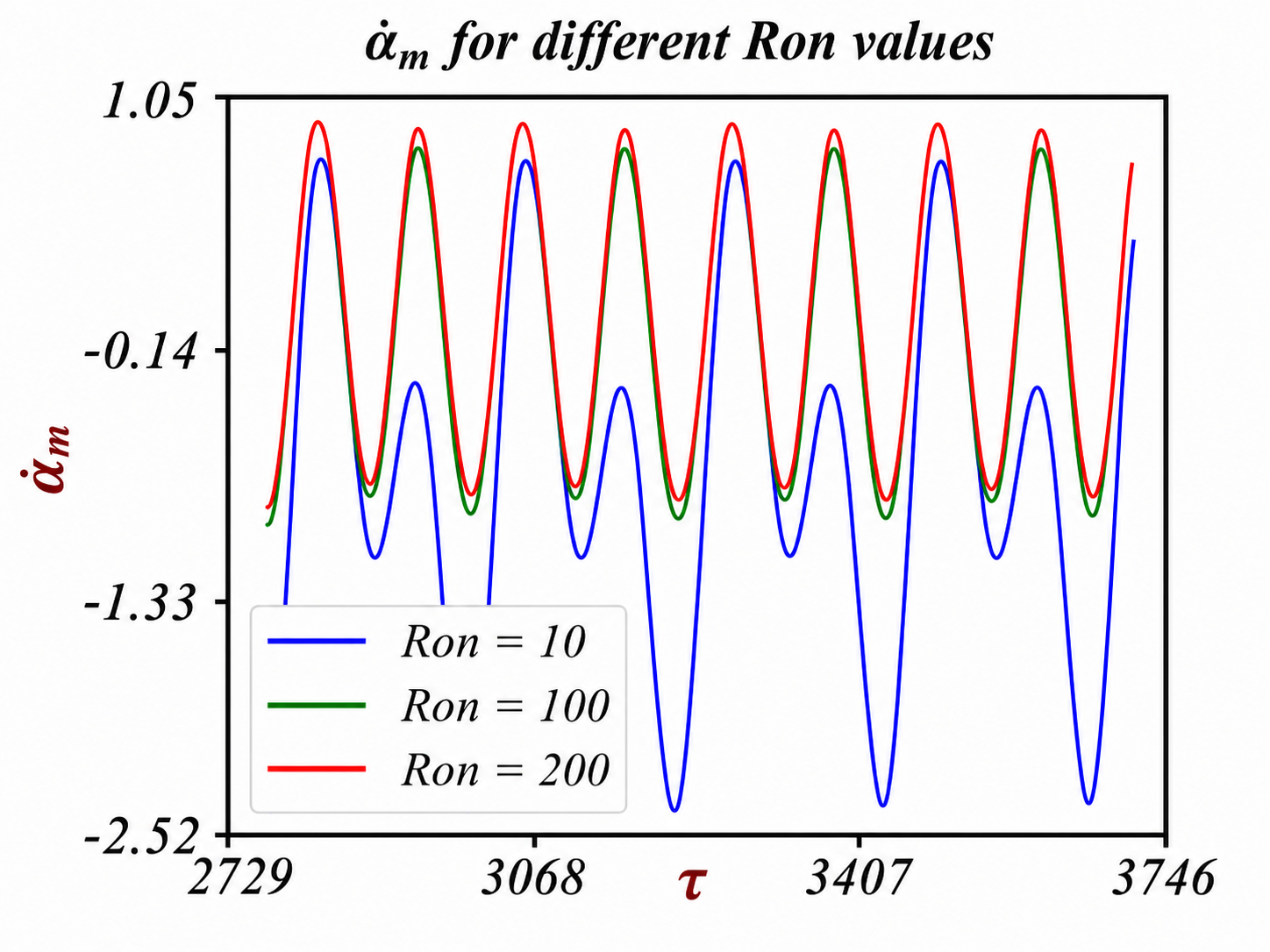}
    \caption{Memristor voltage $\dot{\alpha}_m$ time evolution}
\end{subfigure}
\hfill
\begin{subfigure}{0.48\linewidth}
    \centering
    \includegraphics[width=4.6cm, height=3cm,trim=0cm 0cm 0cm 3cm,clip]{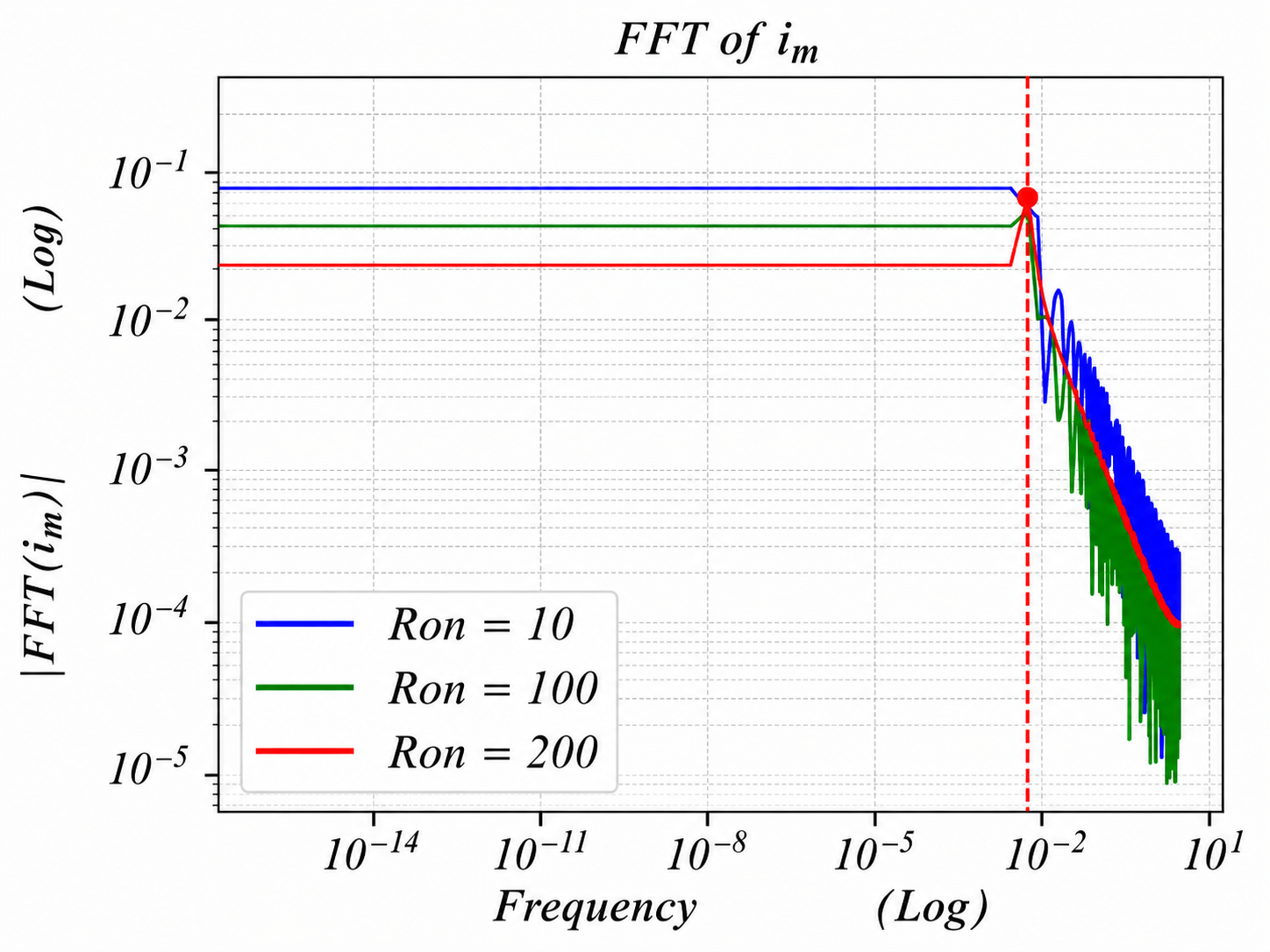}
    \caption{Fourier amplitude spectrum of $i_m$} 
\end{subfigure}
\begin{subfigure}{0.98\linewidth}
    \centering
    \includegraphics[width=12cm, height=3cm,trim=0cm 0cm 0cm 2.4cm,clip]{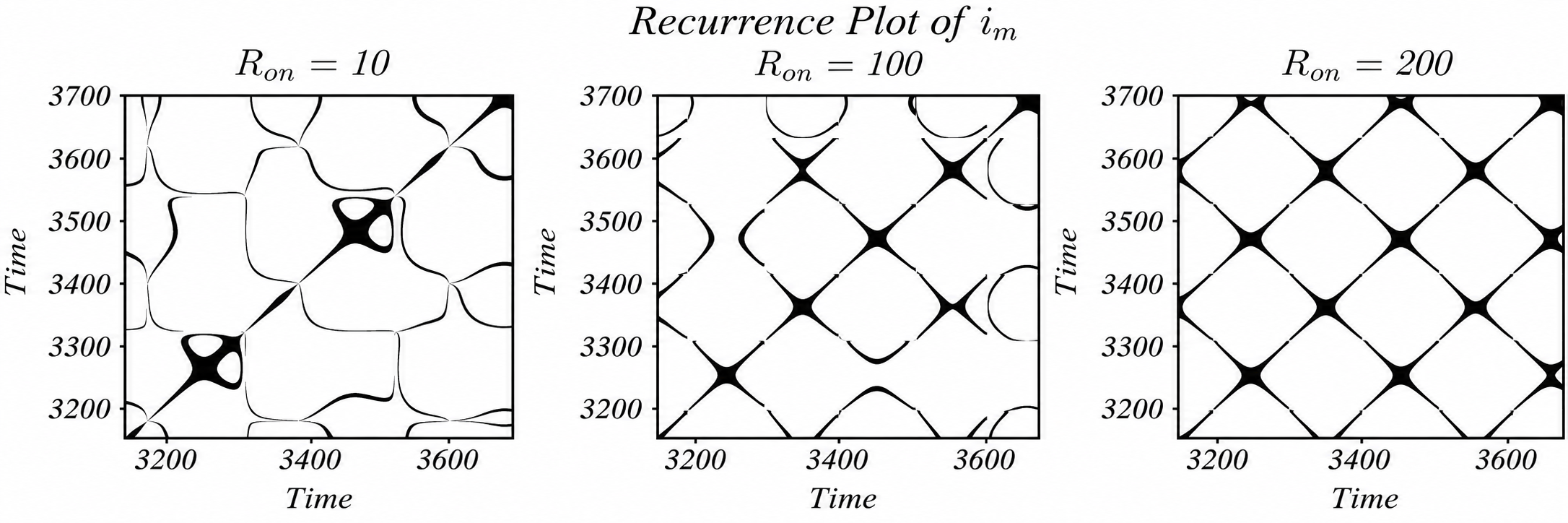}
    \caption{$R_{{on}}$-driven evolution of the recurrence shape of the memristor current $i_m$ \,\,with \,\,$r=0.1$.} 
\end{subfigure}
\caption{$(a)$--$(c)$ and $(e)$:Illustration of quasi-periodic electrical dynamics of the memristor oscillations. $(d)$: Highlighting  the electrical noise for different values of $R_{on}$ and $I_0$ selected within the uniform quasi-periodic region of the maximum LE map shown in Fig.~\ref{fig42}. Parameters: $R_{off}=12\,\,{k}\Omega$, $R_{on}=\textcolor{red}{200},\,\textcolor{green}{100},$ and $\textcolor{blue}{10}\,\Omega$, $D=5\,\,{nm}$, $\mu_v=10^{-10}\,\,{cm^2\,s^{-1}\,V^{-1}}$, and $I_0=2\,\,{mA}$.}
\label{fig6}
\end{figure}

Fig.~\ref{fig6} illustrates the oscillatory regimes of the memristor current $i_m$ through its phase portraits, time evolution, Fourier amplitude spectrum, and recurrence plots for different values of $R_{on}$. The values of $R_{on}$ and $I_0$ were selected within the quasi-periodic region identified in Fig.~\ref{fig42}, thus ensuring that the system operates in a regular oscillatory domain.  Recurrence plots (Fig.~\ref{fig6},(e)) exhibit a dominant main diagonal in all cases, which confirms deterministic behavior. However, a clear evolution in the degree of periodicity 
is observed as $R_{on}$ increases. For $R_{on}=10\,\Omega$, 
the recurrence plot lacks pronounced secondary diagonals, 
indicating a weak periodic structure despite the apparent regularity of the time-domain signal. This behavior is consistent with the Fourier amplitude spectrum of the memristor current $i_m$, shown in Fig. \ref{fig6}(d), which reveals a broadened and noisy frequency distribution, 
suggesting multi-frequency oscillations. As $R_{on}$ increases, the oscillations become progressively more structured. 
Recurrence plots display well-defined secondary diagonals and finer line patterns 
(e.g., for $R_{on}=200\,\Omega$), 
indicating enhanced periodic recurrence. 
The Fourier amplitude spectrum of the memristor current $i_m$ becomes more concentrated around dominant frequencies, 
reflecting a transition towards more coherent and quasi-monochromatic oscillations. These observations demonstrate that increasing $R_{on}$ promotes greater dynamical regularity, 
reducing spectral dispersion and reinforcing the periodic character of the memristor current.

\begin{center}
 \textbf{{\it{Numerical analysis of the current flowing through mechanical and electrical resonators}}}\\
 \end{center}

\begin{figure}[h]
\centering
\begin{subfigure}{0.48\linewidth}
    \centering
    \includegraphics[width=5cm, height=3.5cm,trim=0cm 0cm 0cm 3cm,clip]{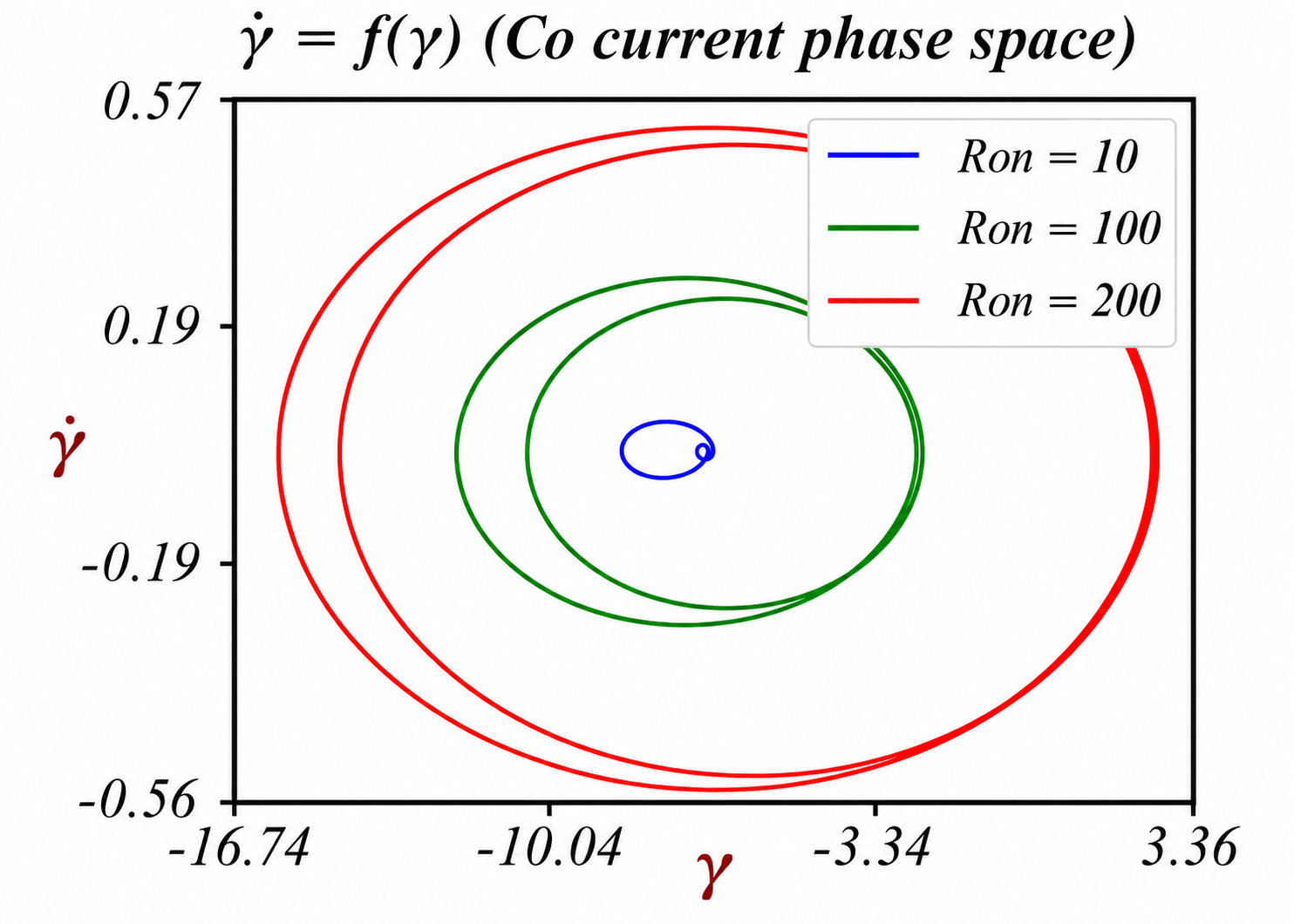}
    \caption{Capacitor $C_0$ charge $\gamma$ phase portrait}
\end{subfigure}
\hfill
\begin{subfigure}{0.48\linewidth}
    \centering
    \includegraphics[width=4.6cm, height=3.5cm,trim=0cm 0cm 0cm 2.2cm,clip]{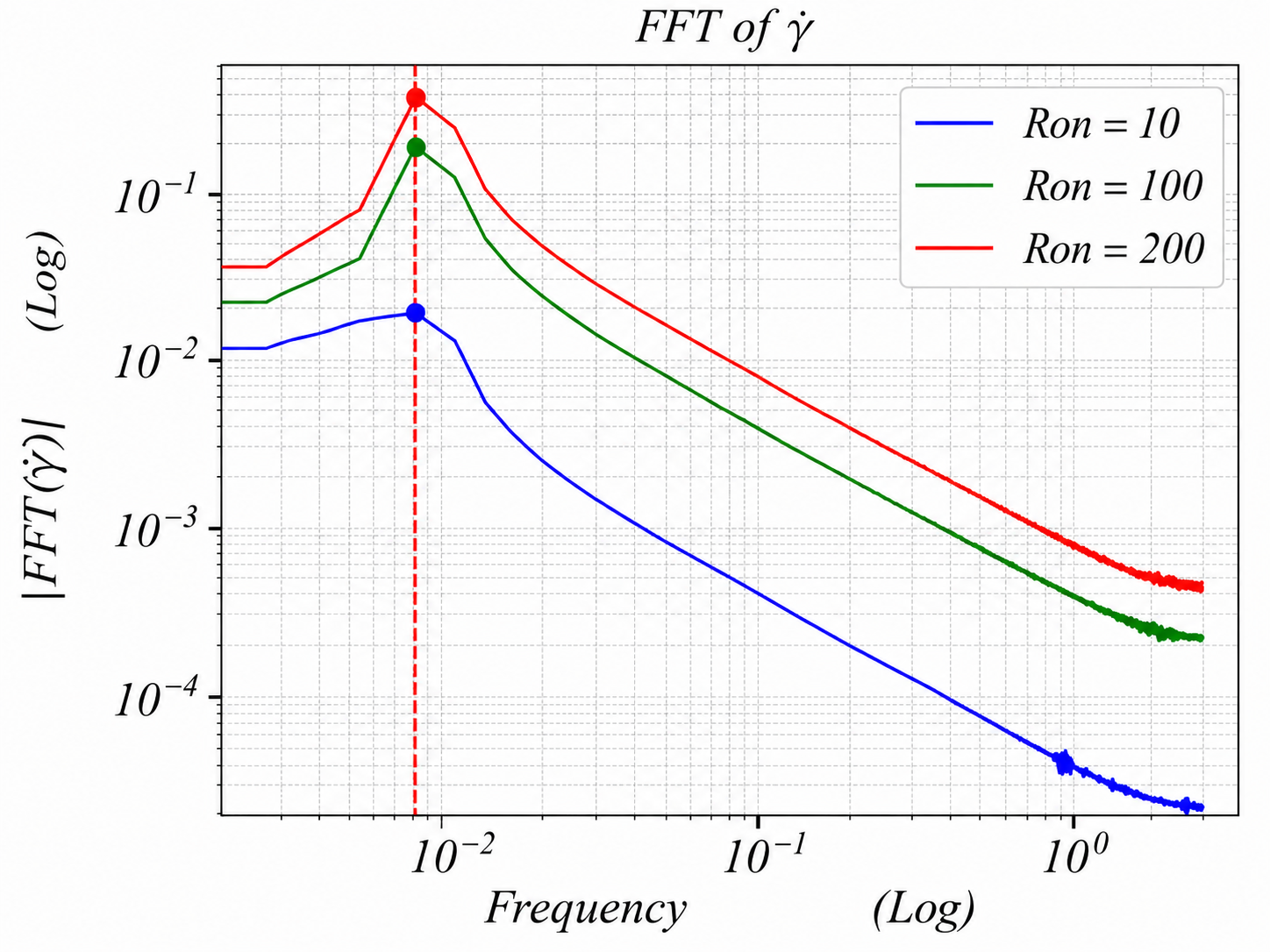}
    \caption{Fourier amplitude spectrum of $i_b$ current.}
\end{subfigure}

\begin{subfigure}{0.98\linewidth}
    \centering
    \includegraphics[width=12cm, height=4cm,trim=0cm 0cm 0cm 2.2cm,clip]{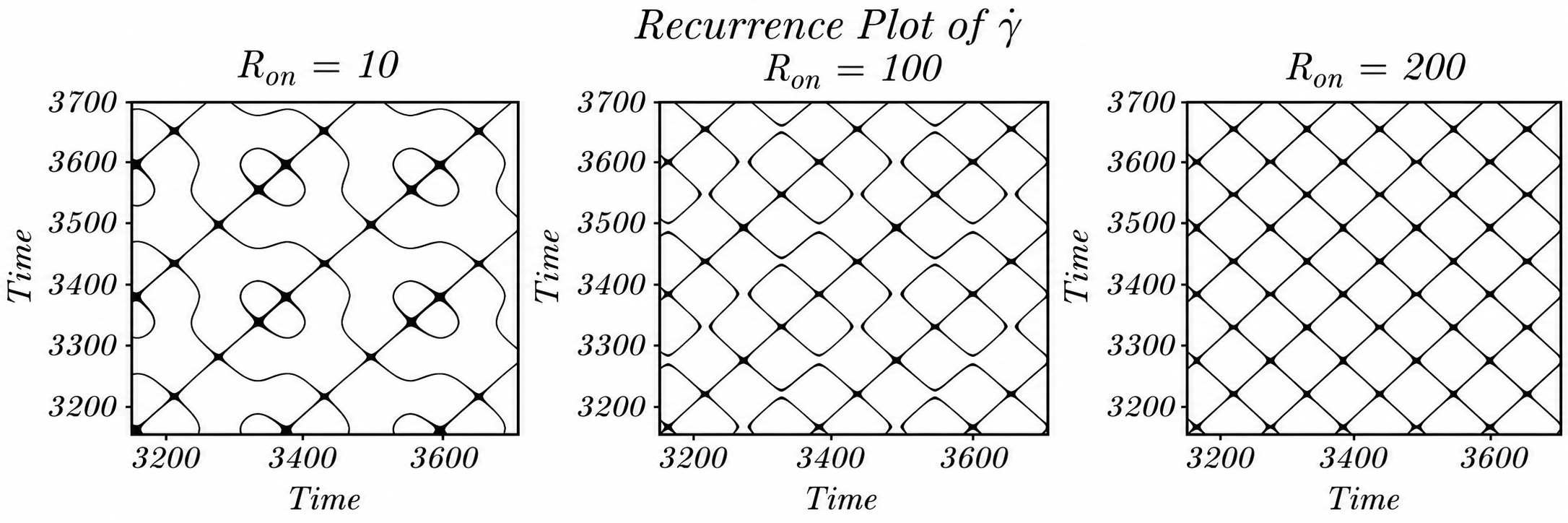}
    \caption{$R_{{on}}$-driven evolution of the recurrence shape of the  capacitor $C_0$ charge current $\dot{\gamma}$}
\end{subfigure}

\caption{\,\,(a), (c): Illustration of the quasi-periodic dynamics  of the capacitor $C_0$ charge current $\dot\gamma \sim i_b$. ($b$): Fourier amplitude spectrum of $i_b$, showing amplitude power gain as $R_{on}$ increases. $R_{on}$ and $I_0$ chosen within the uniform quasi-periodic oscillations region of the  maximum LE map in Fig.~\ref{fig42}. Parameters:
$R_{off} = 12 \,\,{k}\Omega$, 
$R_{on}= \textcolor{red}{200},\, \textcolor{green}{100},$ \,\,and\,\, $\, \textcolor{blue}{10}\,\,\Omega$, \, $D = 5\,\,nm,\,\, \mu_v= 10^{-10}\,\, cm^2 s^{-1}V^{-1}$, and $I_0=2\,\,mA$. }
\label{fig7}
\end{figure}

\begin{figure}[h]
\centering

\begin{subfigure}{0.48\linewidth}
    \centering
    \includegraphics[width=5cm, height=3.5cm,trim=0cm 0cm 0cm 3.2cm,clip]{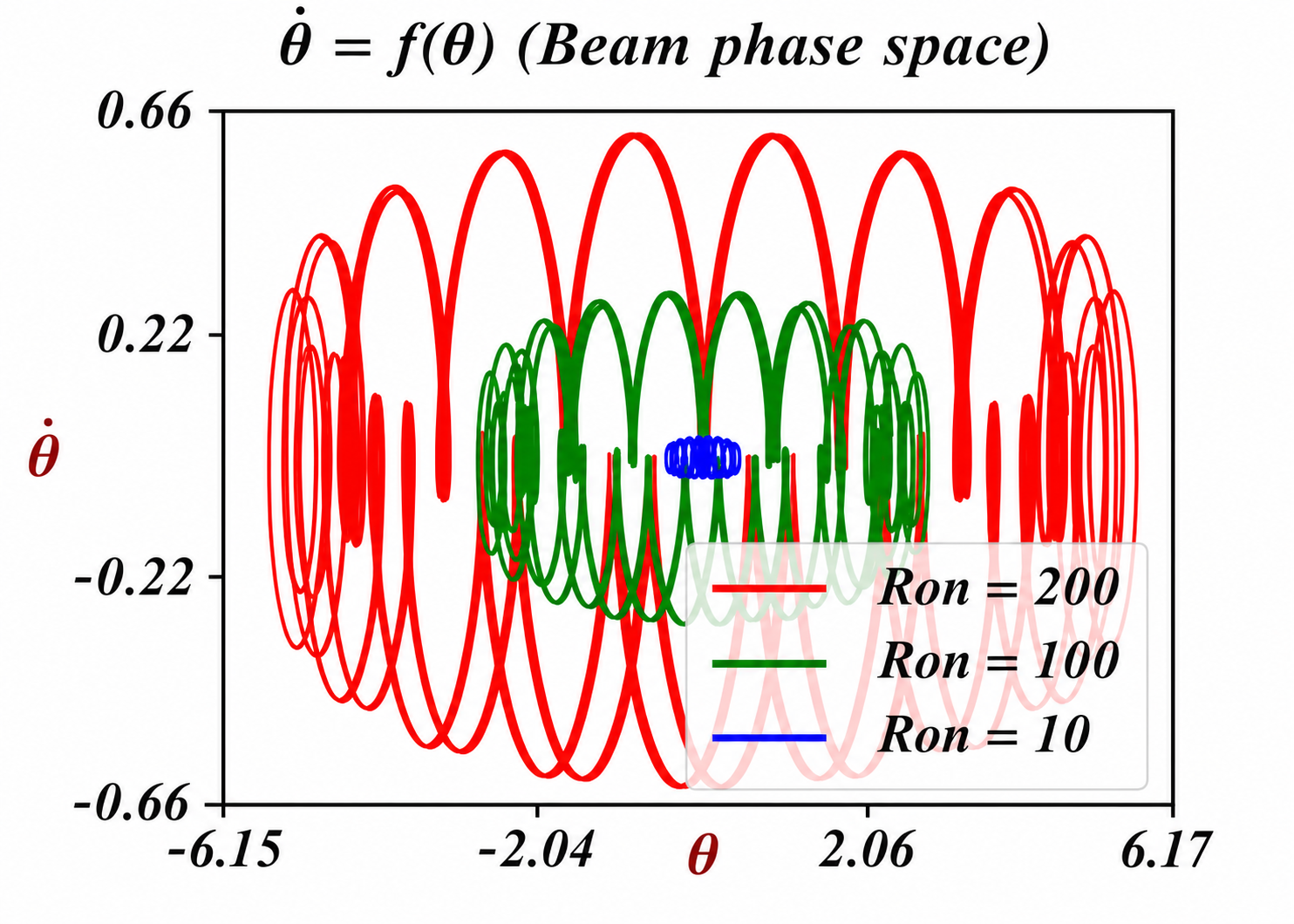}
    \caption{Beam oscillation phase space $\dot{\theta} = f(\theta)$}
\end{subfigure}
\hfill
\begin{subfigure}{0.48\linewidth}
    \centering
    \includegraphics[width=4.cm, height=3.2cm,trim=0cm 0cm 0cm 2.2cm,clip]{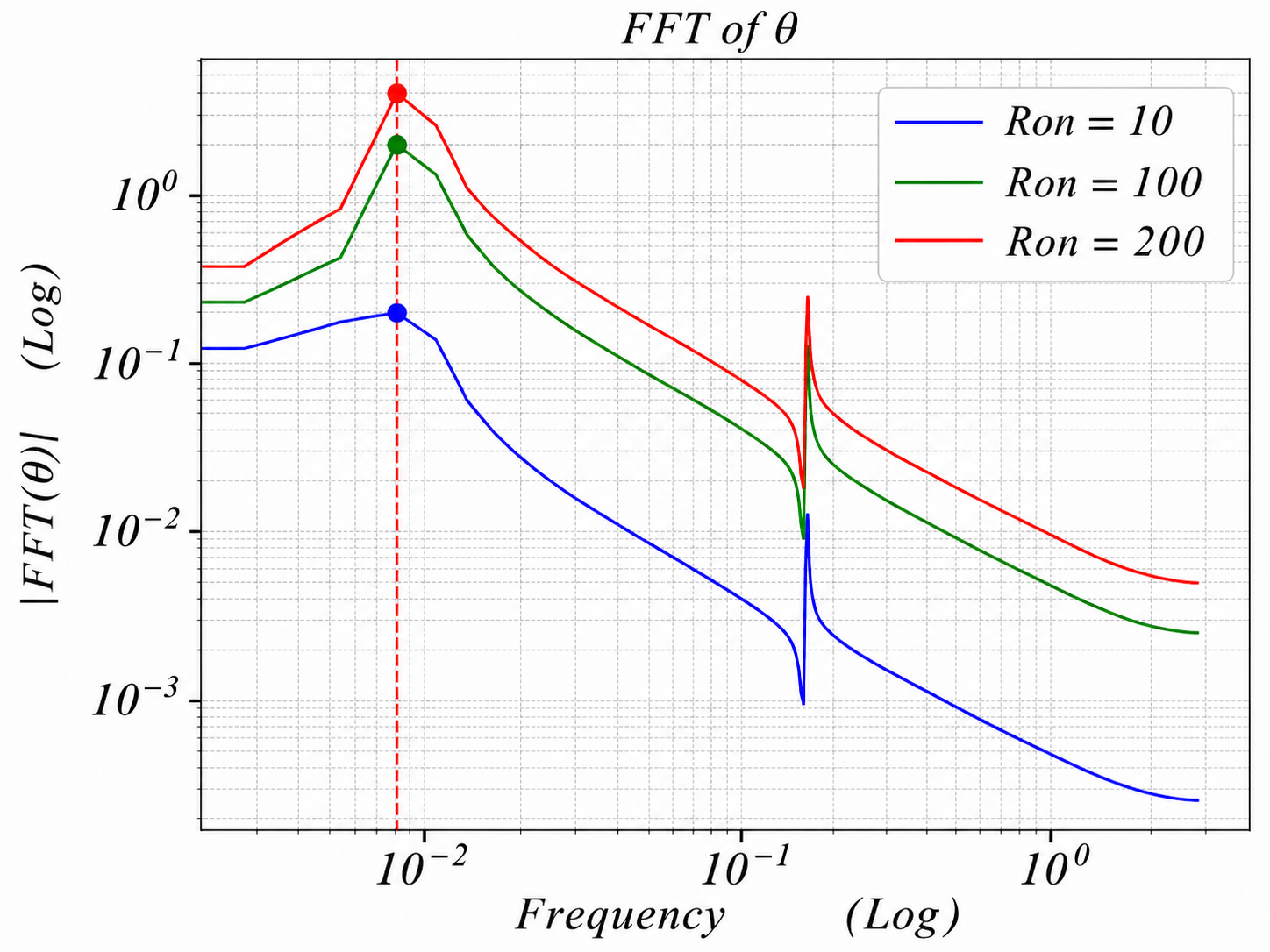}
    \caption{Fourier amplitude spectrum of $\theta$.}
\end{subfigure}

\begin{subfigure}{0.98\linewidth}
    \centering
    \includegraphics[width=12cm, height=4cm,trim=0cm 0cm 0cm 2.4cm,clip]{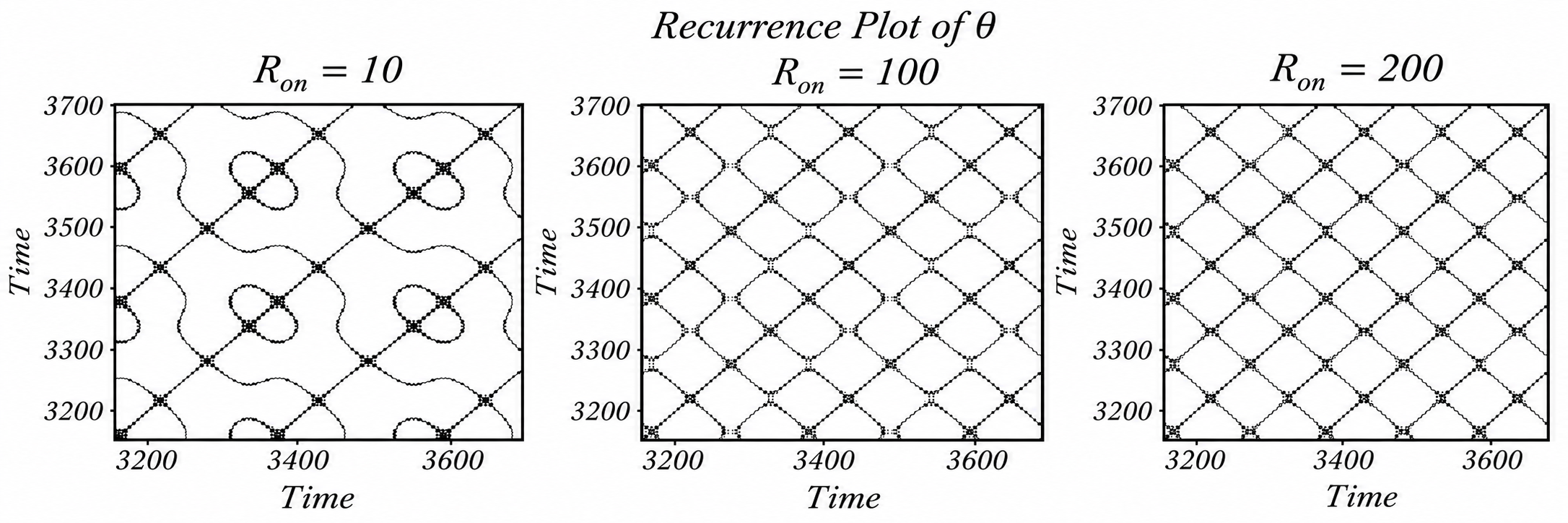}
    \caption{$R_{{on}}$-driven evolution of the recurrence shape of the beam amplitude \,\,$\theta$.}
\end{subfigure}

\caption{
(a),(c): Illustration of the quasi-periodic dynamics of the beam amplitude $\theta$.   
(b): Highlighting the mechanical double resonance frequencies and  power gain as $R_{on}$ increases. 
The parameters $R_{on}$ and $I_0$ are selected within the  uniform quasi-periodic region of the maximum LE map in Fig.~\ref{fig42}. Parameters:
$R_{off} =12\,\,{k}\Omega$, 
$R_{on}= \textcolor{red}{200},\, \textcolor{green}{100},$ \,\,and\,\, $\, \textcolor{blue}{10}\,\,\Omega$, 
$D = 5\,{nm}$, 
$\mu_v = 10^{-10}\,cm^2\,V^{-1}\,s^{-1}$, 
and $I_0 = 2\,\,{mA}$.}
\label{fig8}
\end{figure}

The phase-space portraits shown in Fig.~\ref{fig7} exhibit bi-periodic and highly regular oscillatory trajectories for all values considered  of $R_{on}$. This regularity is further confirmed by the recurrence plots, 
which display well-defined primary and secondary diagonals, 
indicating strong periodic structure and deterministic behavior.
Both the charge $\gamma$ and the charging current $\dot{\gamma} \sim i_b$ increase proportionally with $R_{on}$. This scaling behavior explains the predictable and stable response of the MEMS, which remains consistent with Kirchhoff’s current and voltage laws. As $R_{on}$ increases, the memristor becomes more resistive and consequently draws less current (see Fig.~\ref{fig6}). This redistribution of current results in a larger fraction of the electrical energy being transferred to the electromechanical resonators (see Fig.~\ref{fig7}). The Fourier amplitude spectrum of the $\dot{\gamma} \sim i_b$ current reveals a single dominant resonance frequency 
that remains essentially unchanged with $R_{on}$, confirming that the system operates in a stable resonant regime. Moreover, the spectral content becomes cleaner at higher $R_{on}$ values, with reduced 
high-frequency noise components. This behavior reflects an internal redistribution of electrical power 
within the coupled system: as $R_{on}$ increases, the memristor dissipates 
less electrical power (see  Fig.~\ref{fig6} (d)), while the electromechanical resonators experience a corresponding increase in electrical power (see Figs.~\ref{fig7} and \ref{fig8}). This energy transfer between subsystems remains fully consistent with Kirchhoff’s laws and the global conservation of current and energy.

\begin{center}
 \textbf{{\it{Numerical analysis of  mechanical oscillation of the beam}}}\\
 \end{center}

The recurrence plots of the mechanical oscillations of the beam, 
represented by the transverse displacement $\theta$ of the center of the beam  (see Fig.~\ref{fig8}, (c)), exhibit a dynamical structure that closely resembles that of the charging current $i_b$. The dotted recurrence lines reflect slight modulation effects while preserving deterministic quasi-periodic behavior.
 The corresponding phase-space portraits in Fig.~\ref{fig8}(a) reveal 
quasi-periodic trajectories for all values of $R_{on}$. Here, $\theta$ denotes the transverse deflection amplitude of the center of the beam with respect to its equilibrium position, thus directly characterizing  the mechanical vibration state of the resonator. The maximum deflection amplitude increases proportionally with $R_{on}$, indicating improved electromechanical coupling and greater mechanical energy transfer from the electrical subsystem to the beam. The Fourier amplitude spectrum of the beam amplitude $\theta$ displays two dominant resonance peaks, confirming 
a bi-frequency response consistent with quasi-periodic motion. 
Moreover, the mechanical spectral amplitude power increases with $R_{on}$, 
while no broadband components are observed. The absence of spectral spreading indicates that the mechanical oscillations 
remain coherent and dynamically stable within the explored parameter range. These results demonstrate that increasing $R_{on}$ strengthens the mechanical response of the beam without inducing chaotic  frequency components, thus preserving a structured quasi-periodic regime.

The first numerical  investigations reveal two distinct electromechanical dynamical behaviors  depending on the regions identified in the maximum Lyapunov exponent maps. When the parameters $(R_{on},I_0)$ are selected within the uniform quasi-periodic regions, the coupled MEMS tends to exhibit organized and dynamically stable oscillations (see Figs.~\ref{fig6},~\ref{fig7}, and~\ref{fig8}). In these regimes, the effective memristive resistance $R(t)$
appears to evolve more regularly, which may maintain a nearly uniform electromechanical coupling and promote coherent quasi-periodic responses in both the electrical and mechanical subsystems. In contrast, parameters chosen within the transitional regions of the Lyapunov maps tend to produce a form of dynamical coexistence between quasi-periodic and chaotic oscillations (see Figs.~\ref{fig42} and~\ref{fig91}). Stronger fluctuations of the internal state variable $w(t)$ may induce more irregular variations of the effective resistance $R(t)$, which could continuously modulate the nonlinear electromechanical coupling and contribute to complex energy redistribution between the electrical and mechanical resonators. This mechanism may explain the coexisting chaotic and quasi-periodic dynamical behaviors observed in the temporal responses, bifurcation structures, and Poincar\'e sections, without necessarily implying the  systematic coexistence of distinct attractors.

\subsubsection{Control of chaotic and quasi-periodic regimes through memristor parameters}

\begin{figure*}[!htb]
    \centering
    \begin{subfigure}[b]{0.45\textwidth}
        \centering
        \includegraphics[width=5cm, height=4cm]{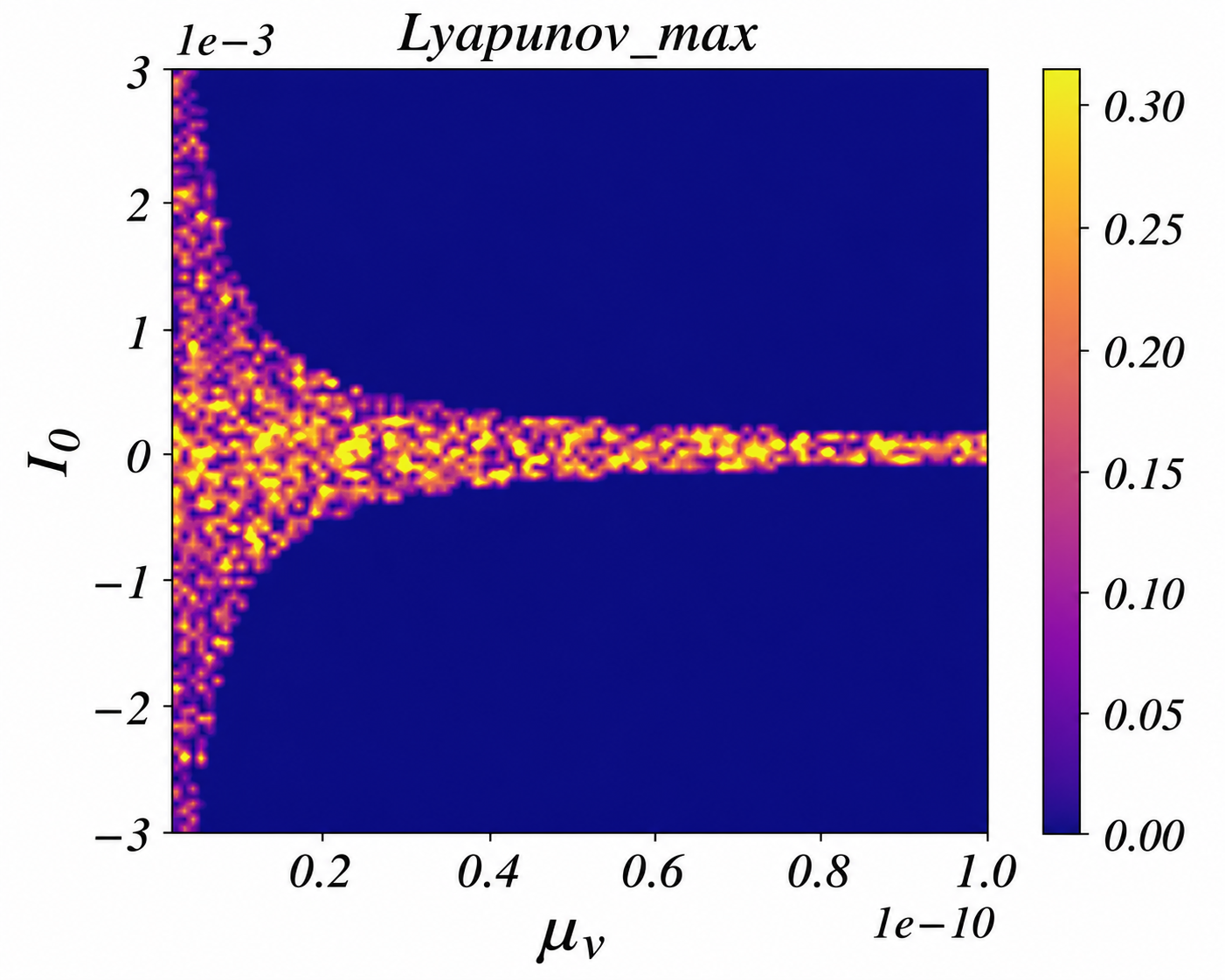}
        \caption{Max LE $ I_0 = f(\mu_v)$ for $D=5nm$ }
        \label{fig:graphe1}
    \end{subfigure}
    \hfill
    \begin{subfigure}[b]{0.45\textwidth}
        \centering
        \includegraphics[width=5cm, height=4cm]{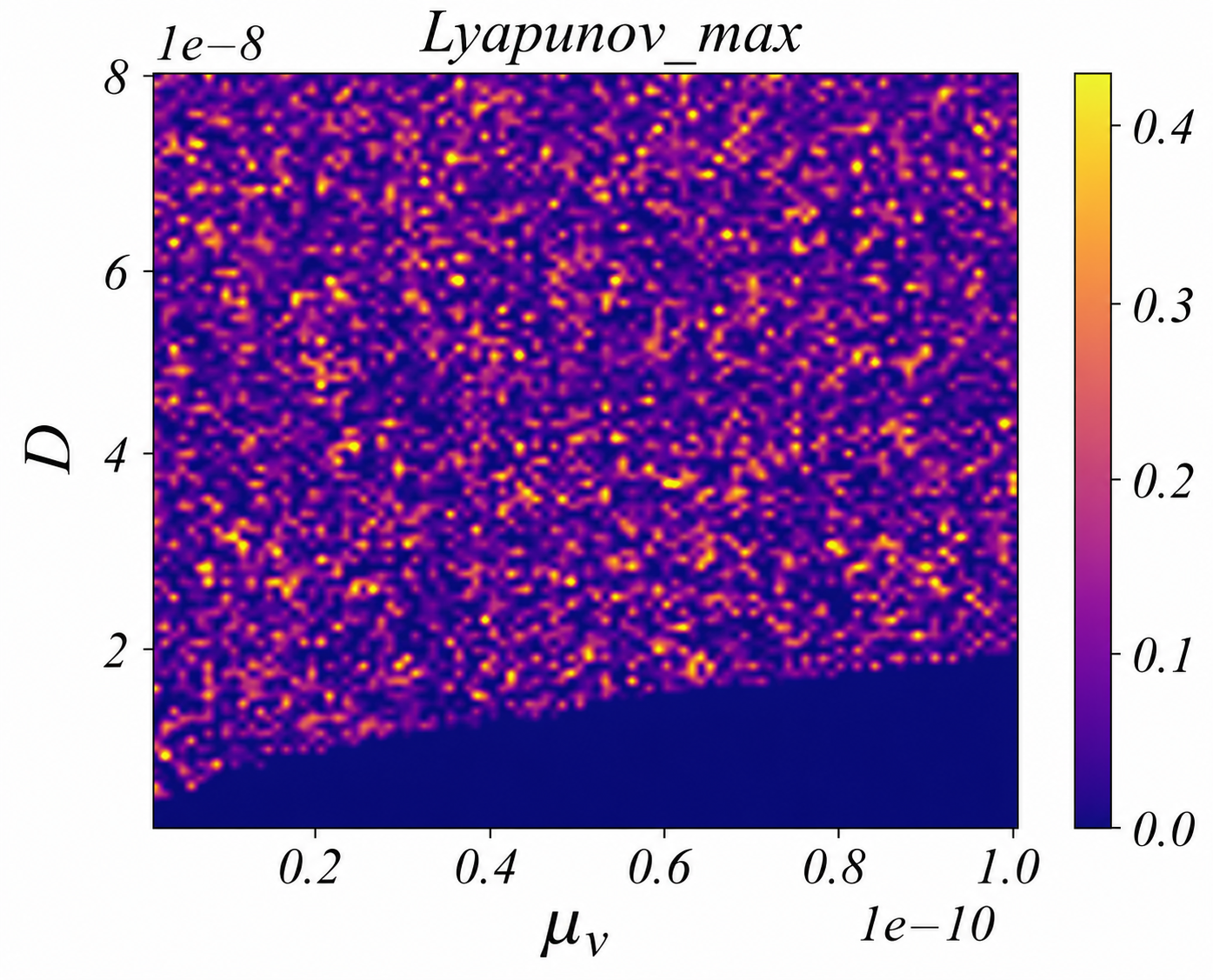}
        \caption{Max LE $D = f(\mu_v)$ for $I_0=-1mA$}
        \label{fig:graphe2}
    \end{subfigure}
    \caption{Influence of the memristor ionic mobility $\mu_V$ and geometric length $D$ on the control of chaotic and quasi-periodic regime boundaries through the maximum Lyapunov exponent. Panels (a) and (b) show the dependence of the max LE on $(\mu_V, I_0)$ and $(\mu_V, D)$, respectively. Here, $D$ denotes the active layer length, while $\mu_V$ represents the ionic mobility. Parameters: $R_{off} = 12k\Omega$,  and $R_{on} = 100\Omega$,\,\, $\mu_v(\mathrm{cm}^2\,\mathrm{V}^{-1}\,\mathrm{s}^{-1})$.}
    \label{fig11}
\end{figure*}

\begin{figure*}[!htb]
    \centering 
    \begin{subfigure}[b]{0.8\textwidth}
        \centering
        \includegraphics[width=12cm, height=6cm]{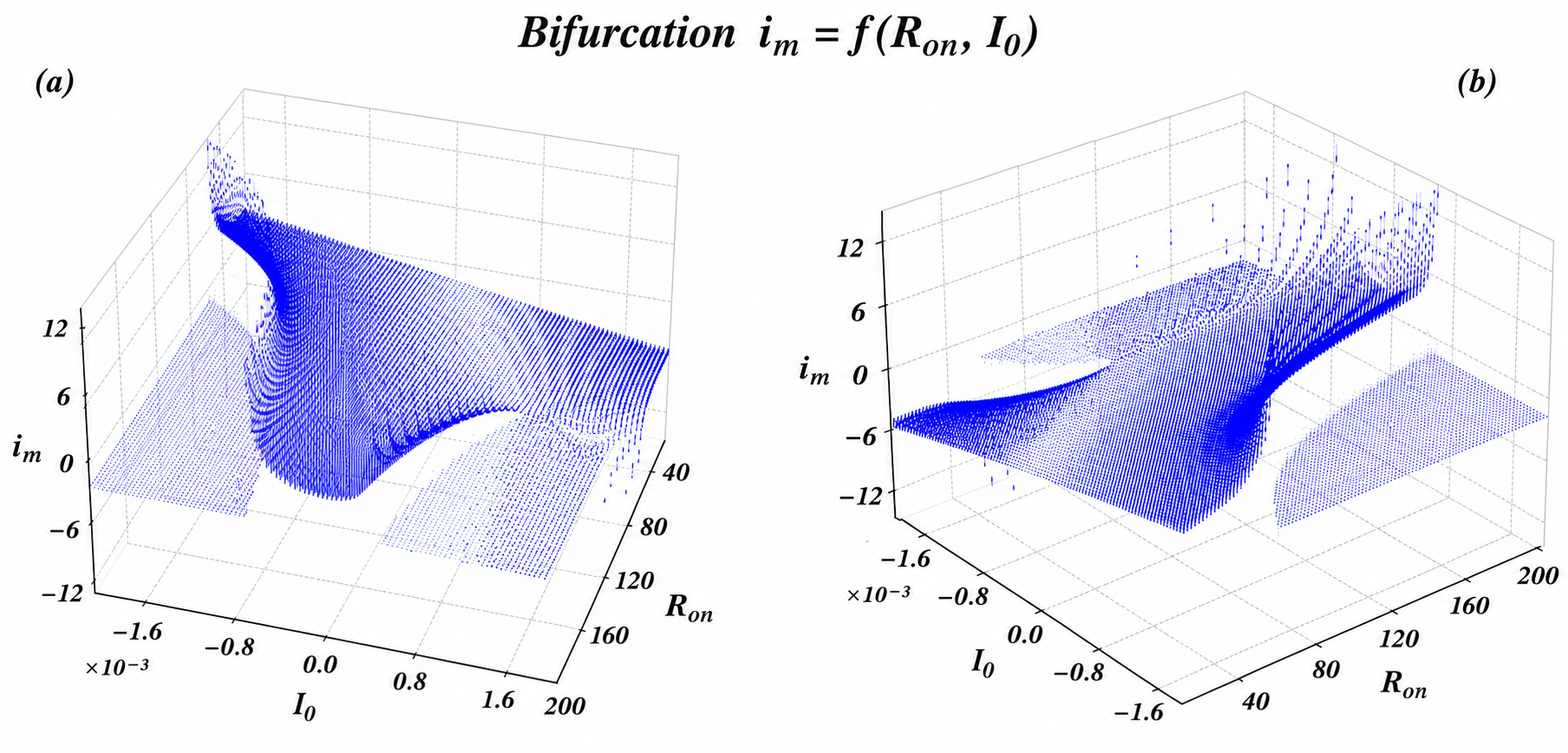}
        \label{fig:graphe1}
    \end{subfigure}
    \begin{subfigure}[b]{0.8\textwidth}
        \centering
        \includegraphics[width=5.5cm, height=4.5cm]{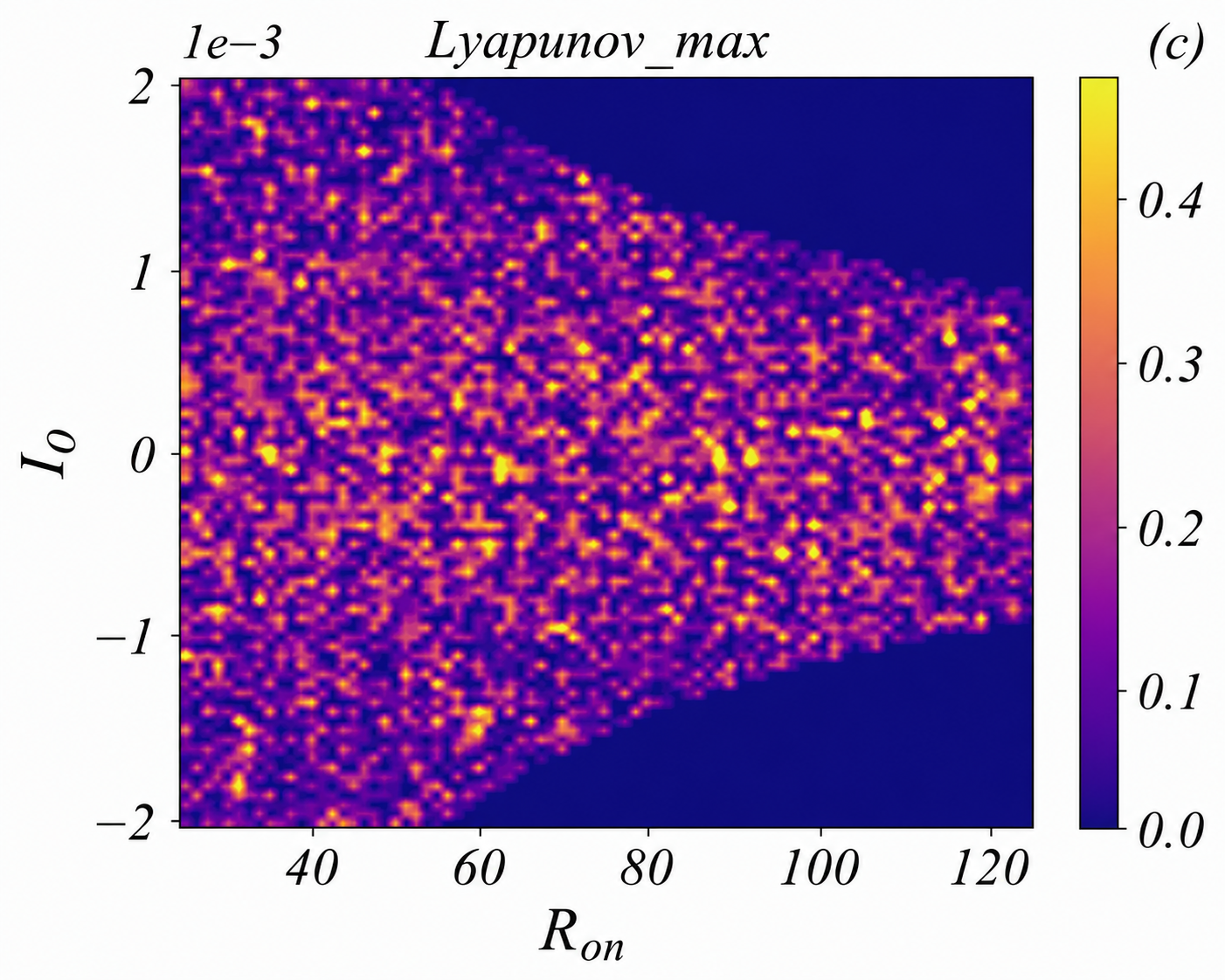}
        \vspace{-0.1cm}
        \label{fig:graphe2}
    \end{subfigure}
    \caption{Bifurcation structures and maximum Lyapunov exponent distributions of the  memristive MEMS as functions of $R_{on}$ and $I_0$ for $\mu_v = 8\times10^{-12}\,{cm^2\,s^{-1}\,V^{-1}}$, $R_{off}=12\,\,{k}\Omega$, and $D=5\,\,{nm}$. Panels (a) and (b) show two viewing orientations of the bifurcation diagram of the memristor current $i_m$, while panel (c) presents the corresponding Lyapunov map highlighting the extension of alternating chaotic and quasi-periodic dynamical regions.}
    \label{fig11.1}
\end{figure*}

\begin{figure}[h]
  \centering
  {\includegraphics[width=9cm, height=3.5cm,trim=0cm 0cm 0cm 0cm,clip]{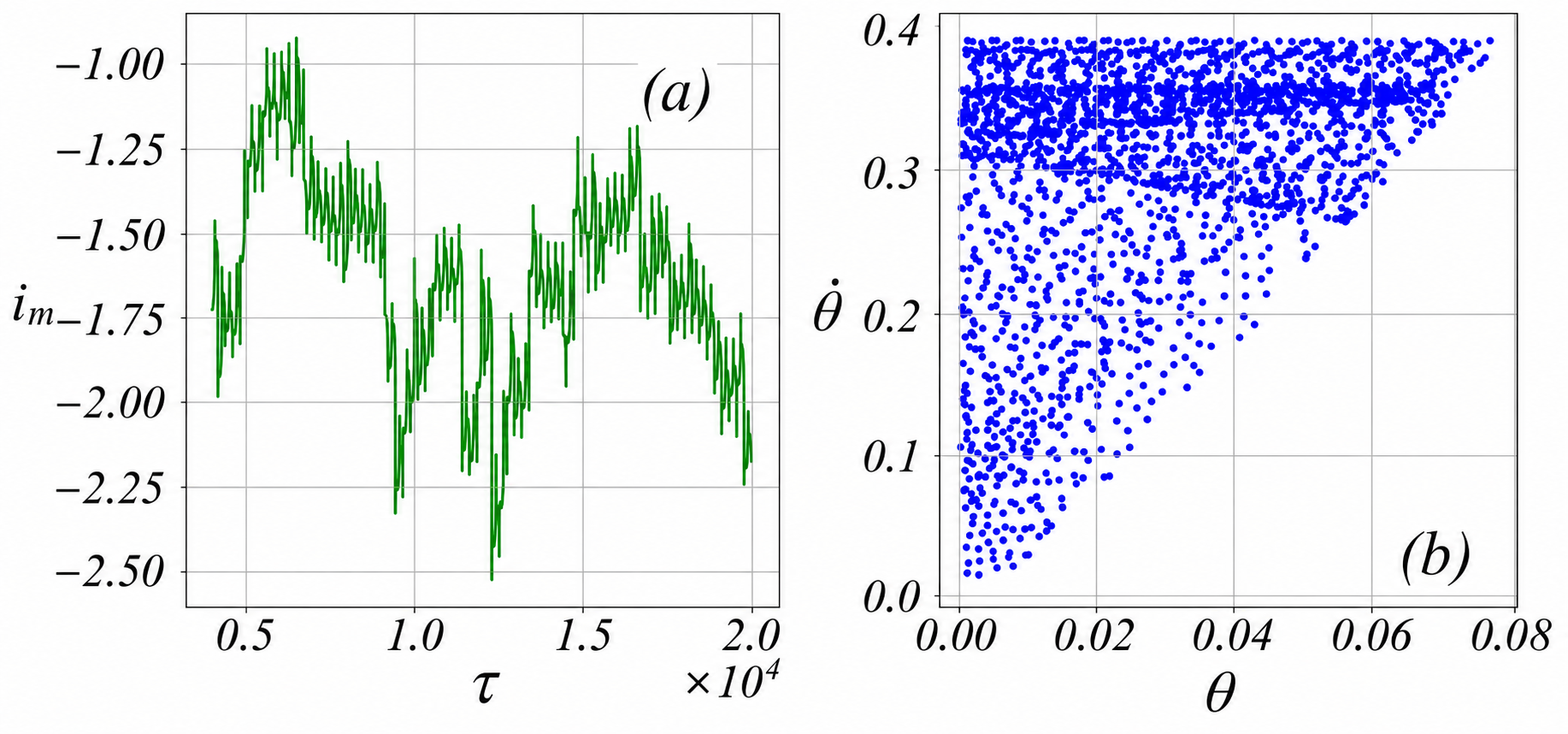}} 
  \caption{Accentuation of the chaotic regime illustrated in Fig.~\ref{fig42} for $\mu_v = 8\times10^{-12}\,\,{cm}^2\,{V}^{-1}\,{s}^{-1}$. 
(a) Time evolution of the memristor current. 
(b) Beam Poincaré section for $D = 5\,\,{nm}$, $I_0 = -1.5\,\,{mA}$, $R_{on} = 60\,\,\Omega$, and $R_{off} = 12\,\,{k}\Omega$.} \label{fig15}
\end{figure}

To highlight the nonlinear dynamics with the memristor parameters, we plot in Fig.\ref{fig11} the maximum values of the Lyapunov exponents (LE) with another parameter of the memristor. As the ion mobility  $\mu_v$ governs the charge transport dynamics of the memristor, we choose to evaluate its influence on the oscillatory regimes of the system when the amplitude $I_0$ of the excitation current varies. Figs. ~\ref{fig11}(a) and ~\ref{fig11}(b) illustrate an amplification of the strictly positive LE domain as $\mu_v$ decreases.
For $\mu_v = 8\times10^{-12}\,\,cm^2 s^{-1}V^{-1}$, Fig.~\ref{fig11.1}, shows an expansion of the chaotic domain of $LE=f(R_{on}, I_0)$ obtained in Fig.~\ref{fig42}. While the chaotic zone was almost zero for $R_{on} \ge 50\Omega$ in Fig.~\ref{fig42}, the diagram in Fig.~\ref{fig11.1} predicts chaotic regimes for the values of $R_{on} \ge 100\,\,\Omega$ and $\lvert I_0 \rvert \ge 1\,\,mA$. The bifurcation diagram of Fig.~\ref{fig11.1}(a) \,\,and\,\,~\ref{fig11.1}(b) confirms a sharp reduction in regions of pure quasi-periodicity in favor of an extended positive Lyapunov exponent domain.
  The more accentuated chaotic behavior can be seen for $R_{on} = 60\,\Omega, I_0 = -1.5\,\,mA$  in Fig.~\ref{fig15}. Despite a value of $R_{on} = 60\,\Omega$ three times higher than that chosen to obtain the chaotic dynamics in Fig.~\ref{fig91}, Fig.~\ref{fig15} highlights a marked reinforcement of the chaotic regime in the memristor current, accompanied by a pronounced dispersion in the Poincaré section of the mechanical oscillations. As the beam deflection 
amplitude decreases corresponding to an increase in the memristor current 
(see Figs.~\ref{fig6} and ~\ref{fig7}), the mechanical trajectories become 
progressively more irregular. In other words, the intensification of chaotic 
oscillations in the memristor current induces a loss of regularity in the 
mechanical response of the beam. This behavior reflects the strong nonlinear coupling between the electrical and mechanical subsystems: when the memristor current enters a strongly 
chaotic regime, the mechanical oscillations inherit this irregularity 
through the electromechanical interaction.

\subsection{Thermo-induced bifurcation dynamics of the memristive MEMS}
\par
\vspace{1cm}

\qquad Temperature plays a crucial role in the electrical behavior of memristive devices. In the HP memristor model, the total resistance depends on the relative position of the boundary between the conductive doped region $TiO_{2-x}$ and the insulating region $TiO_2$ \cite{strukov}. In realistic experimental memristive devices, indirect coupling between ionic mobility, conductivity, and resistive states may arise through self-heating effects, local defect redistribution, or thermally activated vacancy migration mechanisms~\cite{Ielmini2016}. Several experimental and theoretical studies have demonstrated that the principal memristive parameters $R_{on}$, $R_{off}$, and $\mu_v$ exhibit significant temperature sensitivity~\cite{singh2018temperature,garcia2016spice,Khan2024,abunahla2016modeling}. Consequently, temperature-dependent formulations reported in the literature are considered in the present work  to qualitatively interpret the thermo-sensitive nonlinear dynamics of the proposed MEMS architecture. Based on experimentally supported conductivity relations, the ionic mobility of oxygen vacancies and the resistive states of the HP memristor may be expressed as ~\cite{singh2018temperature}:

\begin{equation}
\mu_v =
\frac{\sigma \, M(\mathrm{TiO_2})}
{N_A \, e \, n \, \rho(\mathrm{TiO_2})}
\label{eq:mobility}
\end{equation}

\begin{equation}
R_{{on}} =
\frac{w}{\sigma(T) A_m}
\qquad \text{and} \qquad
R_{{off}} =
\frac{D-w}{\sigma(T) A_m}.
\label{eq.Ron}
\end{equation}

The electrical conductivity $\sigma(T)$ follows the temperature-dependent relation

\begin{equation}
\sigma(T)=
\sigma_{0,\mathrm{Mott}} T_{p}^{-2s}
\exp\left[
-\left(
\frac{T_{0,\mathrm{Mott}}}{T_p}
\right)^s
\right]
+
\sigma_{0,\mathrm{ES}} T_p^{-2s}
\exp\left[
-\left(
\frac{T_{0,\mathrm{ES}}}{T_p}
\right)^s
\right].
\label{eq:conductivity}
\end{equation}

which combines both Mott variable-range hopping and Efros--Shklovskii hopping conduction mechanisms frequently reported in oxide-based memristive materials~\cite{singh2018temperature,Khan2024}. $T_p$ is the temperature, and the other parameters are fully described  in ~\cite{singh2018temperature}. These thermo-activated relations indicate that moderate temperature variations may substantially modify the effective ionic transport and the internal resistive state of the memristor, thereby affecting the nonlinear electrodynamic response of the coupled thermo-electro-mechanical MEMS. For completeness, the temperature-dependent evolutions of $\mu_v$, $R_{on}$, and $R_{off}$ reported by Singh and Raj ~\cite{singh2018temperature} are illustrated in Fig.~\ref{fig14} in the  Appendix. In the present work, temperature effects are assumed to act globally only on the memristive parameters in order to avoid additional perturbations associated with the strong thermo-mechanical hypersensitivity of the microbeam. Nevertheless, local self-heating effects in the HP TiO$_2$ memristor may induce non-uniform thermal gradients capable of locally modifying the ionic mobility $\mu_v$, the ON-state resistance $R_{{on}}$, and the OFF-state resistance $R_{{off}}$.
 As the temperature increases, resistances $R_{on}$ and $R_{off}$ tend to decrease due to the increase in electronic conductivity, while ionic mobility $\mu_v$ increases as a result of the thermally activated diffusion mechanisms.

\begin{figure*}[!htb]
    \begin{subfigure}[b]{0.35\textwidth}
        \centering
        \includegraphics[width=6cm, height=4.8cm]{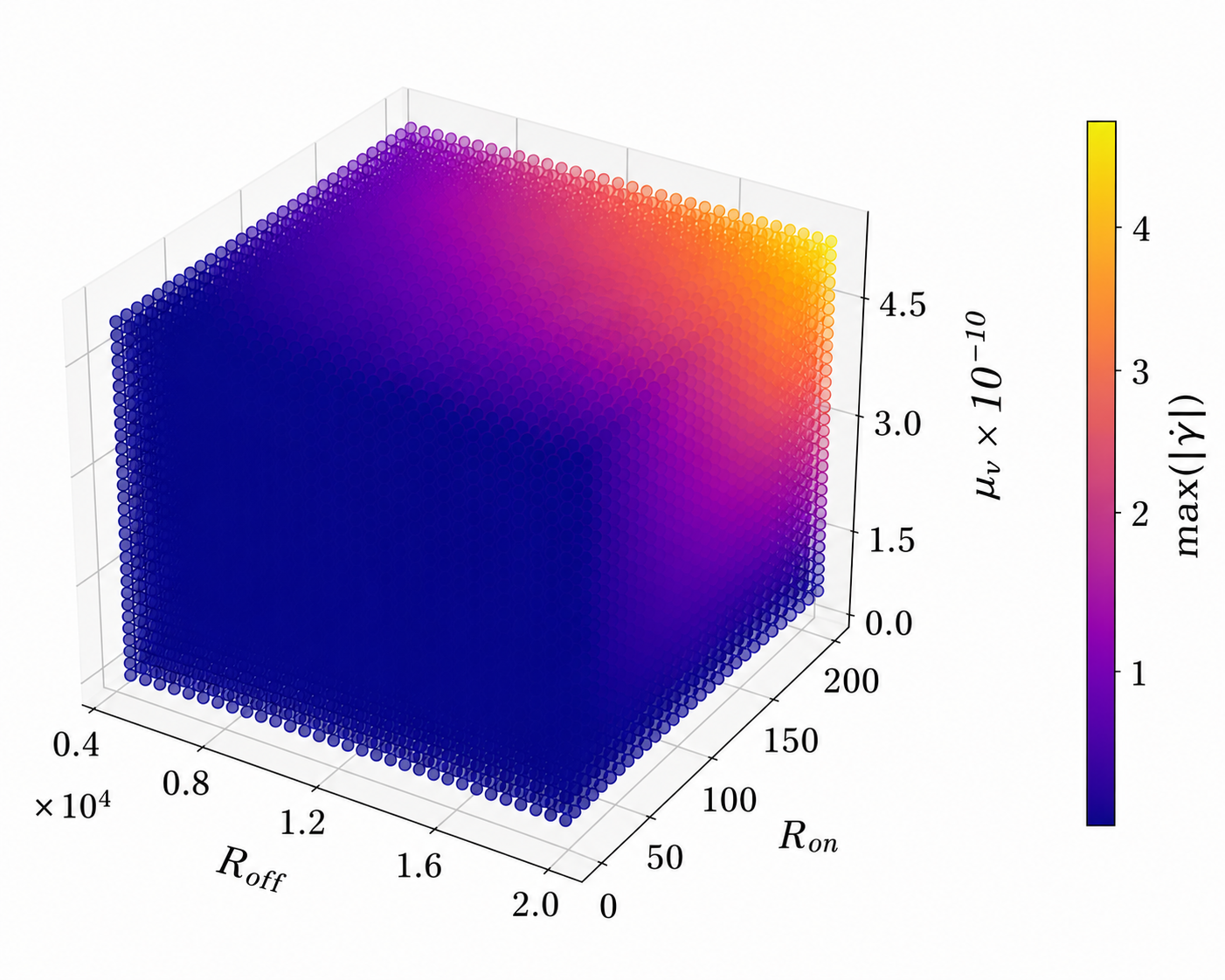}
        \subcaption{$\dot{\gamma} = f(R_{on},\,\, R_{off},\,\, \mu_v)$: Response in the current $\dot{\gamma} \sim i_b$ flowing through the microbeam as a function of the thermo-sensitive parameters of the HP memristor.}
        \label{fig:graphe1}
    \end{subfigure}
    \hspace{2.5cm}
    \begin{subfigure}[b]{0.35\textwidth}
        \centering
        \includegraphics[width=6cm, height=5.2cm]{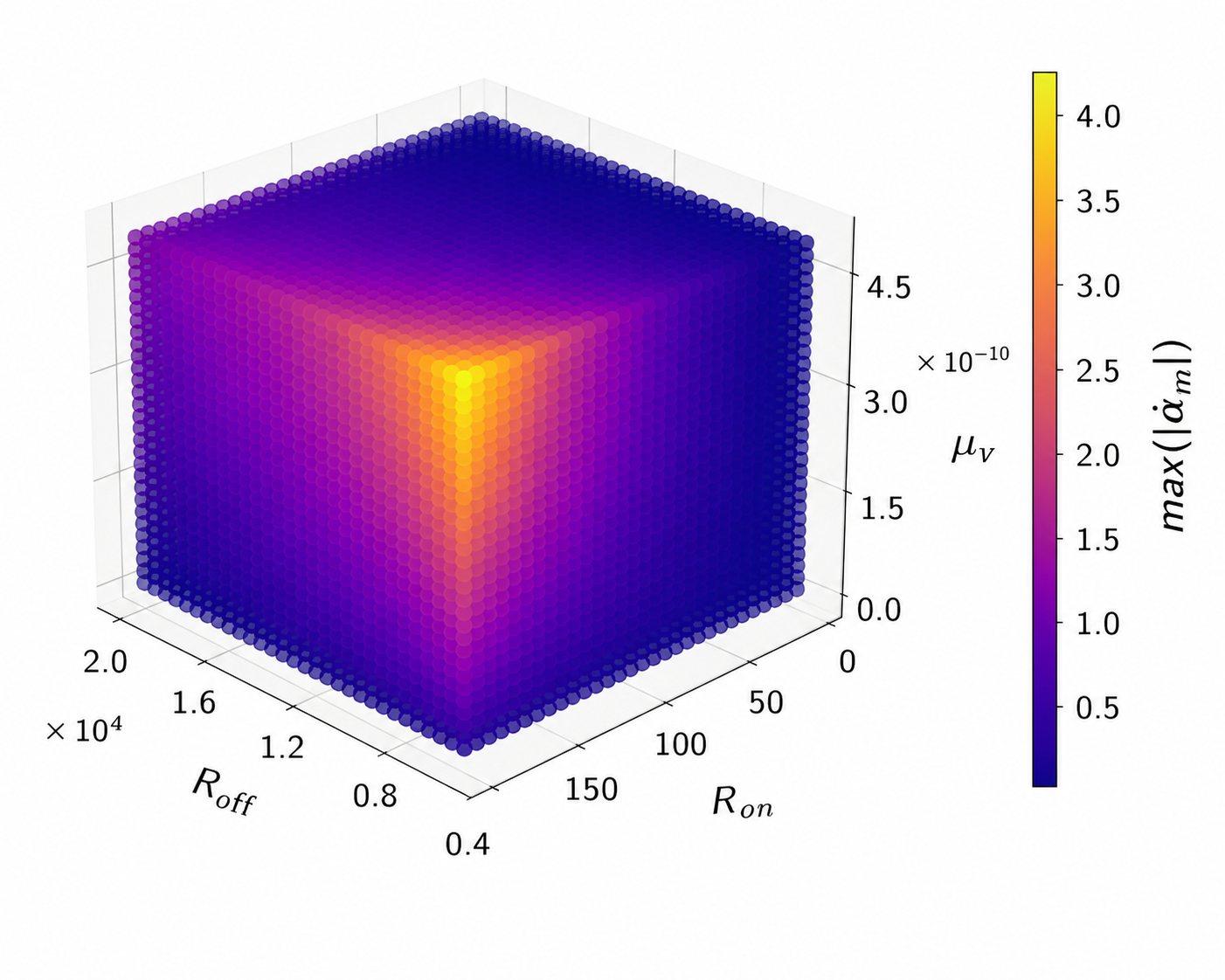}
        \subcaption{$\dot{\alpha}_m = f(R_{on},\,\, R_{off},\,\, \mu_v)$: Voltage $\dot{\alpha}_m \sim V_m$ across the memristor  as a function of the thermo-sensitive parameters of the HP memristor.}
        \label{fig:graphe2}
    \end{subfigure}
    \caption{4D diagram illustrating: (a) the sensitivity of the current $i_b \sim \dot{\gamma}$ flowing through the mechanical (micro-beam) and electrical $(r_0,L_0,C_0)$ resonators and (b) the sensitivity of the voltage $V_m \sim \dot{\alpha}_m$ across the memristor, as a function of the three thermo-sensitive parameters $(R_{on},\,\, R_{off} ,\,\,\mu_v)$ of the HP memristor. Parameters: $\mu_v \,(cm^2 s^{-1}V^{-1})$,  $R_{on}\,(\Omega)$, \,\, $R_{off} \,(\Omega)$, $I_0 = 2\,\,mA$,\,\, and \,\,$D=5\,\,nm$.} \label{fig4D}
\end{figure*}

Figure~\ref{fig4D} was generated by simultaneously varying these three ($R_{on},\,\, R_{off},\,\, \mu_v$) HP memristor parameters. It clearly shows in Fig. \ref{fig4D} (a), the current response $i_b \sim \dot{\gamma}$, represented here by the fourth color dimension (color bar), which propagates through both the mechanical (micro-beam) and electrical $(r_0, L_0, C_0)$ resonators. A strong dependence of the electrical current amplitude $i_b$ is observed on the  memristor parameters $R_{on}$, $R_{off}$, and $\mu_v$.
Since the deflection of the micro-beam is proportional to the current intensity $i_b$, it can be concluded that both the mechanical and electrical response amplitudes are thermally dependent on the HP memristor parameters. Indeed, a simultaneous decrease in $R_{on}$, $R_{off}$, and an increase in $\mu_v$ induces a variation of the electrical response amplitudes across the two resonators (see Fig.~\ref{fig4D} (a)). 
Figure~\ref{fig4D}(b) illustrates the variation of the voltage across the memristor. The results reveal a stronger sensitivity to variations of $\mu_v$ and $R_{on}$ than to those of $R_{off}$. Indeed, only slight modifications of $\dot{\alpha}_m$ and $\dot{\gamma}$ are observed when $R_{off}$ decreases, whereas significantly larger variations occur for changes in $R_{on}$ and $\mu_v$. These observations suggest that the thermo-induced energy redistribution within the coupled electromechanical system remains weakly sensitive to variations of $R_{off}$. Furthermore, the combined analysis of Figs.~\ref{fig4D}(a) and \ref{fig4D}(b) indicates that the thermo-electromechanical dynamics of the system may be more sensitive to variations in the ionic mobility $\mu_v$ and the ON-state resistance $R_{on}$ than to changes in the OFF-state resistance $R_{off}$.

\begin{figure*}[htbp]
    \begin{subfigure}[b]{0.45\textwidth}
        \centering
        \includegraphics[width=6cm,height=5cm,
        trim=0cm 0cm 0cm 7cm,clip]{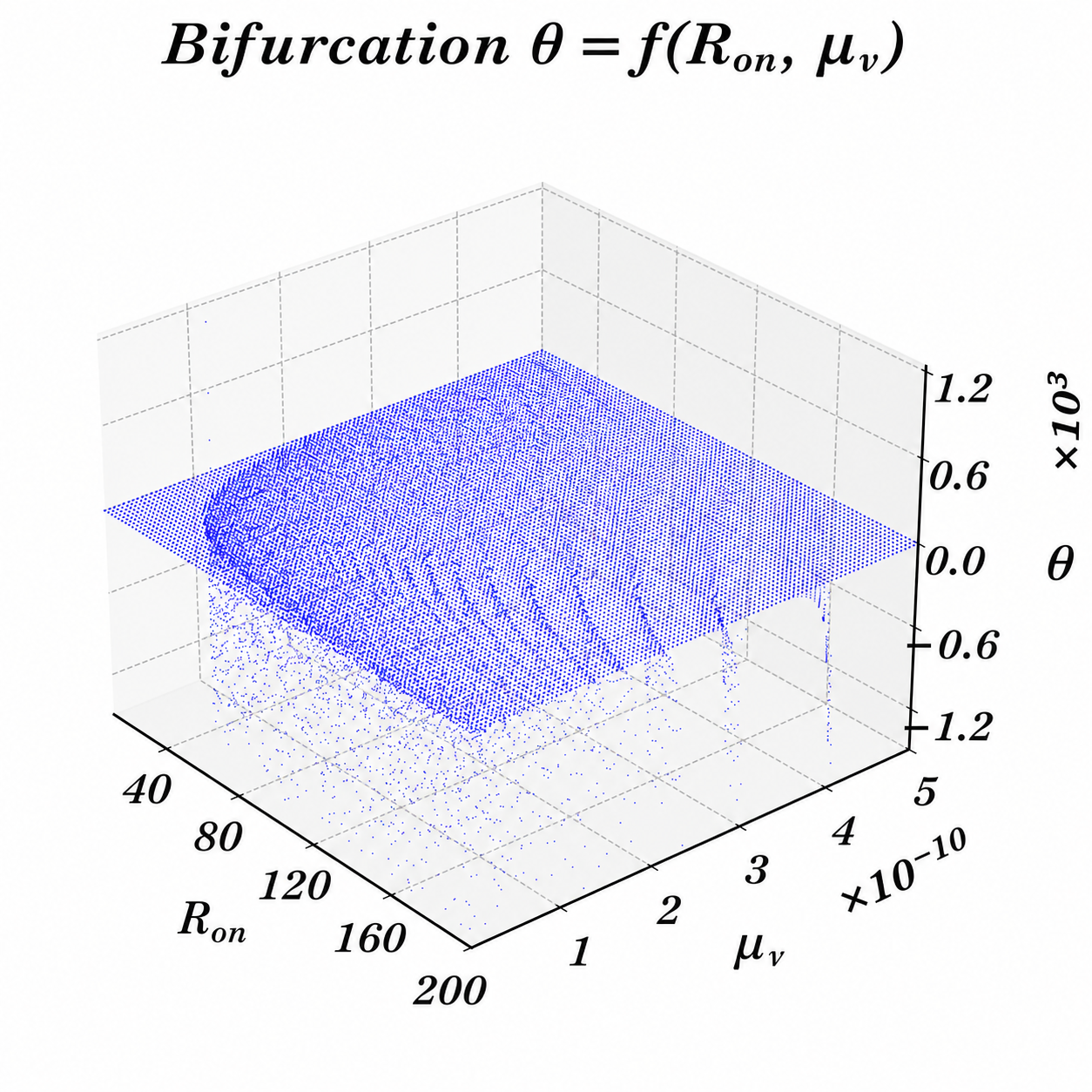}
        \caption{Bifurcation structure of the microbeam amplitude $\theta$ under thermo-electric memristive effects}
        \label{fig9a}
    \end{subfigure}
    \hfill
    \begin{subfigure}[b]{0.45\textwidth}
        \centering
        \includegraphics[width=6cm,height=5cm,
        trim=0cm 0cm 0cm 7cm,clip]{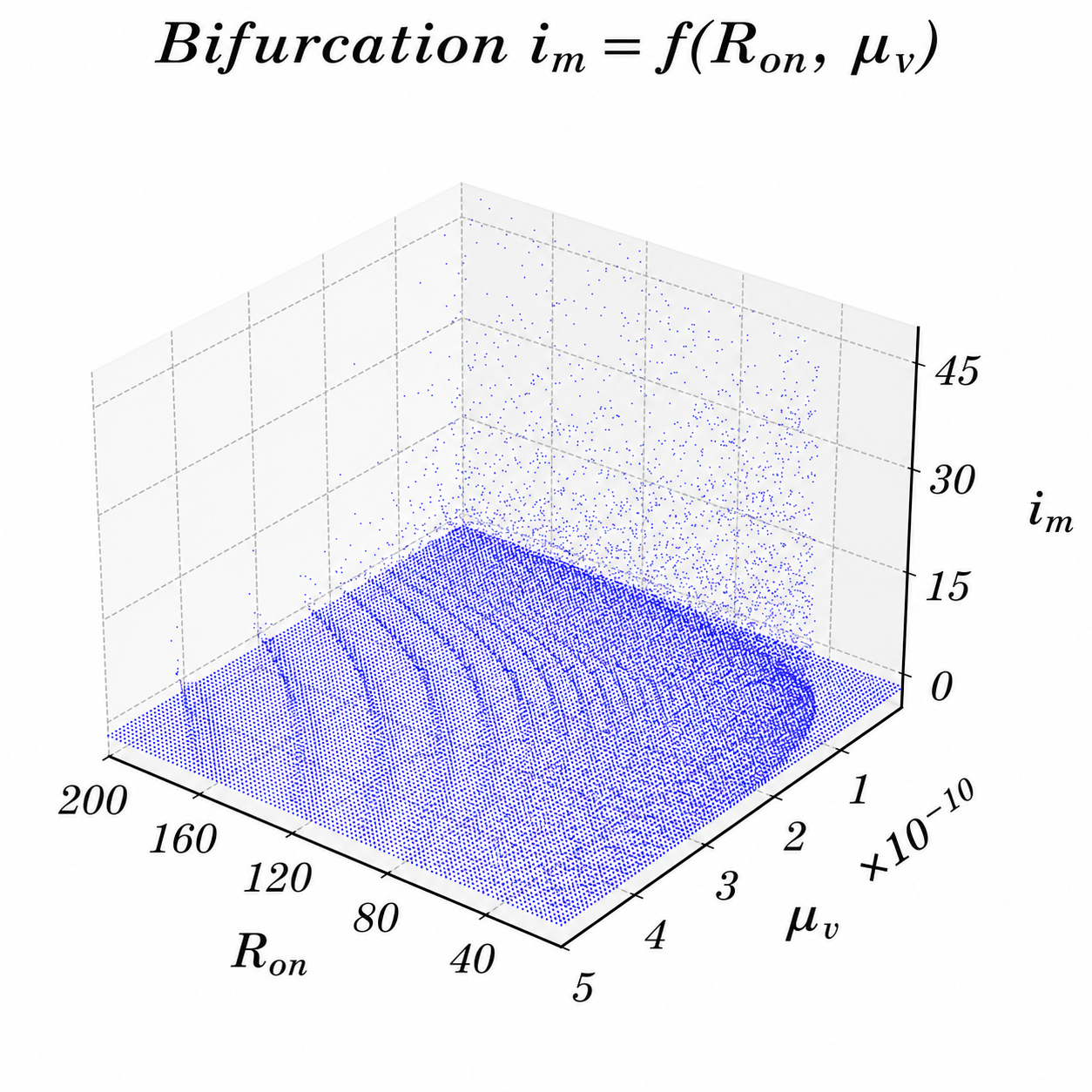}
        \caption{Bifurcation structure of the memristor current $i_m$ under thermo-electric memristive effects}
        \label{fig9d}
    \end{subfigure}

\caption{Cascade of thermo-induced nonlinear electromechanical bifurcation regimes driven by variations of the resistive ratio adjusted through $R_{on}$, and the ionic mobility $\mu_v$. The parameter ranges are $\mu_v \in [10^{-12},\,5.10^{-10}]\,{cm}^2\,{V}^{-1}\,{s}^{-1}$ and $R_{on} \in [5,\,200]\,\Omega$, $I_0 = -1.5\,{mA}$, and $D = 5\,{nm}$.}
\label{fig92}
\end{figure*}

To investigate the electromechanical responses of the thermo-controlled MEMS to the thermal sensitivity of the HP memristor, the ratio $\dfrac{R_{off}}{R_{on}}$ was varied through modifications of $R_{on}$. As previously shown in Fig.~\ref{fig4}, a range of $R_{off}$ values exists for which the current $i_b$ exhibits an approximately linear evolution, together with a weak sensitivity of the sensibility to thermo-induced energy redistribution shown in Fig.~\ref{fig4D}. Within this parameter range, variations of $R_{off}$ produce only slight amplitude changes without significantly modifying the electrodynamic regimes of the resonators. Consequently, tuning $R_{on}$ provides an efficient way to control the ratio $\dfrac{R_{off}}{R_{on}}$ and analyze the resulting nonlinear dynamical responses. Figure~\ref{fig92} presents the three-dimensional bifurcation structures associated with the mechanical response of the microbeam and the electrical response of the memristor.  These structures are obtained when the ratio $\dfrac{R_{off}}{R_{on}}$ (adjusted through variations of $R_{on}$) and the ionic mobility $\mu_v$ are simultaneously varied.

Figure~\ref{fig92} illustrates dynamical transitions between 
quasi-periodic and chaotic regimes as a function of the main memristor parameters. 
Both  mechanical and electrical oscillations exhibit a clear tendency toward more pronounced chaotic behavior as ionic mobility $\mu_v$ decreases. The bifurcation structures shown in Fig.~\ref{fig92} illustrate a cascade of thermo-induced dynamical transitions generated by simultaneous variations of the ionic mobility $\mu_v$ and the ON-state resistance $R_{on}$. These transitions suggest the emergence of a thermally controllable temporal coexistence between quasi-periodic and chaotic electromechanical dynamical behaviors. Indeed, Fig.~\ref{fig92} reveals an acceleration of the transitions between chaotic and quasi-periodic modes  when  $\mu_v$ increase and $R_{on}$ decreases. From Figs.~\ref{fig4D} and ~\ref{fig92}, it can therefore be concluded that moderate temperature variations in the HP memristor can control not only the electromechanical oscillation amplitudes (Fig.~\ref{fig4D}) but also the dynamical modes of the system (Fig.~\ref{fig92}). In this sense, temperature acts as an indirect parameter, enabling bifurcation control of  coupled electromechanical dynamics. The dominant quasi-periodic behaviors observed for increasing values of  $\mu_v$, corresponding to an increase in the HP memristor temperature, suggest potential applications of the system as a thermal sensor or thermal actuator within well-defined narrow temperature ranges. However, it should be noted that no large operating region completely excludes the occurrence of chaotic regimes, unlike the case observed in the absence of thermal effects (Fig.~\ref{fig42}). A persistent alternating of chaotic and quasi-periodic regimes is observed, although the frequency of alternation between these regimes decreases significantly as  $\mu_v$ increases  (Fig.~\ref{fig92}(a)).

The numerical results suggest that variations in the intrinsic memristor parameters $R_{on}$, $R_{off}$, $D$, and $\mu_v$ significantly influence the dynamical regimes of the coupled electromechanical system. Such parameter variations enables a rich spectrum of quasi-periodic and chaotic behaviors and allows controlled transitions between these regimes. In particular, since the ionic mobility $\mu_v$, $R_{on}$, and $R_{off}$ are strongly temperature dependent, moderate thermal variations in the vicinity of the memristor may induce a transition  from quasi-periodic to chaotic oscillations.\\
\par

\section{Comparative  analysis:  HP memristor versus Josephson junction dynamical systems}

In a previous study, we investigated the nonlinear coupling between a Josephson junction, a micro-beam, and an electrical resonator within a thermo-electro-mechanical MEMS framework~\cite{Koudaf2}. In the present work, a comparative analysis is performed between that superconducting architecture and the proposed HP memristor---micro-beam configuration in order to highlight their respective operational capabilities, physical constraints, and nonlinear dynamical behaviors in the context of thermo-electro-mechanical system applications.

A major advantage of the HP memristor resides in its ability to operate under standard ambient-temperature conditions, unlike Josephson junctions, which generally require cryogenic environments to maintain superconductivity. This room-temperature operability considerably broadens the practical applicability of the proposed MEMS architecture and facilitates its integration into realistic thermo-sensitive electromechanical systems without the technological constraints associated with superconducting devices. Beyond its room-temperature operability, the HP memristor also offers remarkable flexibility for nonlinear dynamical modelling owing to the strong dependence of its behavior on both intrinsic and geometrical parameters. In particular, the memristive dynamics can be efficiently tuned through the ionic mobility $\mu_v$, the resistive states $R_{on}$ and $R_{off}$, the active layer thickness $D$, and the evolution of the internal state variable $w(t)$. Such parameters directly influence the nonlinear energy transfer mechanisms within the coupled thermo-electro-mechanical system and therefore govern the emergence of distinct oscillatory regimes, including periodic, quasi-periodic, and chaotic states. Moderate temperature variations may significantly modify the ionic transport properties and the effective resistive state of the memristor, thereby affecting the amplitudes, resonance conditions, bifurcation structures, and Lyapunov-bassed dynamics characterization  of the MEMS. Consequently, the memristor naturally acts as a thermo-sensitive nonlinear regulator capable of dynamically redistributing energy between the electrical and mechanical resonators without requiring external bias reconfiguration. Compared with the Josephson-based configuration, the proposed memristive architecture provides greater controllability of nonlinear oscillatory modes over a wider operational temperature range. A concise comparison between the principal characteristics of the HP memristor model and the Josephson junction model is presented in Table~\ref{tab:comparison_JJ_eng}.

\begin{table*}[h]
\centering
\caption{Comparative analysis of the HP memristor model and the Josephson junction model for thermo-electro-mechanical MEMS applications.}
\label{tab:comparison_JJ_eng}

\begin{tabular}{p{2cm} p{6.0cm} p{6.0cm}}
\toprule

\textbf{Aspect} &
\textbf{HP Memristor Model} &
\textbf{Josephson Junction Model} \\

\midrule

Operating Temperature
&
Room-temperature operation; typically investigated within $200$--$450$\,K in TiO$_2$ devices
\cite{singh2018temperature,Vaidya2021a,Vaidya2021b}
&
Cryogenic operation; Nb/Al junctions typically operate around $4.2$\,K, while higher-$T_c$ variants still require cryogenic environments; qubit-related JJ systems often explore mK \, to few K regimes
\cite{BaronePaterno1982,Likharev1986,Tinkham1996,ClarkeBraginski2004}
\\

\midrule

Nonlinear Dynamics Control
&
Strong amplitude modulation; control through $R_{on}$, $R_{off}$, $D$, and $\mu_v$; efficient nonlinear regime transitions (quasi-periodic $\leftrightarrow$ chaotic); enhanced bifurcation sensitivity and Lyapunov-controlled dynamics (see Figs.~\ref{fig42}---\ref{fig15}),
\cite{Chua2014,Ginoux2021,Pham2022}
&
Nonlinear RCSJ dynamics; control through bias current, magnetic flux, and microwave excitations; Shapiro steps; frequency locking; chaotic regimes achievable under strong external driving
\cite{BaronePaterno1982,Likharev1986,Tinkham1996,Kautz1996}
\\

\midrule

Intrinsic Memory Properties
&
Native nonvolatile memory; pinched hysteresis loops; state-dependent resistance; dynamic resistive switching (see Figs. ~\ref{fig6}---\ref{fig8}), 
\cite{strukov,Chua2014}
&
No intrinsic nonvolatile resistive memory; memory achievable through flux storage in superconducting loops (SQUID); hysteretic I--V characteristics possible in underdamped regimes
\cite{ClarkeBraginski2004,BaronePaterno1982,Tinkham1996}
\\

\midrule

Thermo-sensitive Effects
&
Thermally activated ionic transport; temperature-dependent conductivity; vacancy-migration sensitivity; thermo-electro-mechanical coupling; temperature-induced dynamical reconfiguration (see Fig. ~\ref{fig4D}),
\cite{Khan2024,singh2018temperature,Vaidya2021a,Vaidya2021b}
&
Critical current $I_c(T)$ decreases with temperature according to the Ambegaokar--Baratoff relation; thermal fluctuations may induce phase diffusion, self-heating effects, and superconducting instability
\cite{AmbegaokarBaratoff1963,Tinkham1996,Kautz1996}
\\

\midrule

Limits and Key Notes
&
Advantages in memory, low-power operation, and room-temperature adaptability; limitations mainly related to material variability, defect redistribution, and sensitivity of the nonlinear response to thermal and structural fluctuations (see Fig. \ref{fig92}), 
\cite{Khan2024,garcia2016spice}
&
Advantages in nonlinear superconducting circuits and ultra-sensitive sensing; limitations include cryogenic requirements, thermal noise, energy dissipation, and the absence of intrinsic resistive memory
\cite{Tinkham1996,Likharev1986}
\\

\bottomrule
\end{tabular}
\end{table*}

 The memristor behaves not only as a nonlinear memory element but also as an internal thermal transducer capable of sensing and dynamically regulating the entire electromechanical system. Furthermore, the coexistence of homogeneous, strongly nonlinear, and chaotic oscillatory regimes achievable through memristive parameter modulation provides enhanced adaptability compared with the Josephson-based approach. Owing to its room-temperature operation, intrinsic memory effect, reduced mathematical complexity, and strong thermo-electro-mechanical coupling capability, the HP memristor appears to be a highly promising candidate for adaptive MEMS architectures dedicated to nonlinear sensing, thermal monitoring, energy transduction, and adaptive electromechanical control.

\section{Conclusions}
The present work was devoted to the mathematical modeling and numerical investigation of a thermo-controlled memristive MEMS based on the coupling between an HP memristor and electromechanical resonators. The results suggest that the oscillation amplitudes, resonance frequencies, electrical noise, and nonlinear dynamical responses of the coupled electrical and mechanical subsystems may be significantly influenced by intrinsic memristor parameters defined at the design stage. In particular, the global dynamics of the system were shown to depend strongly on the memristive parameters $R_{on}$, $R_{off}$, $D$, $\mu_v$. The numerical investigations revealed transitions between quasi-periodic and chaotic electromechanical regimes together with thermally tunable regions of temporal dynamical coexistence between quasi-periodic and chaotic behaviors. The temporal evolution of the effective memristive state was found to continuously modulate the nonlinear electromechanical coupling, thereby influencing the energy redistribution between the electrical and mechanical resonators. In some operating regimes, the quasi-periodic  behavior of the HP memristor appeared to promote relatively stable thermo-electrical coupling conditions, potentially allowing moderate thermal perturbations to affect the electromechanical responses of the MEMS. The results also indicate that the HP memristor may behave not only as a tunable nonlinear dynamical element capable of generating complex oscillatory responses, but also as a thermo-electro-mechanical transduction component. Chaotic electromechanical oscillations similar to those commonly reported in Josephson-junction-based systems were also observed in the proposed memristive MEMS model over a range of thermally sensitive conditions. Unlike Josephson junctions, which generally require cryogenic operating environments, the HP memristor remains compatible with room-temperature operation, potentially offering greater flexibility for nonlinear MEMS applications involving thermo-sensitive and chaos-assisted dynamical functionalities. Overall, these results suggest that thermo-controlled memristive MEMS may offer promising avenues open new prespectives for adaptive nonlinear oscillators, thermo-sensitive sensing devices, and complex electromechanical dynamical systems.

Although the present work primarily highlights thermally tunable coexistence between quasi-periodic and chaotic electromechanical dynamics, the existence of globally coexisting attractors was not rigorously established within the present numerical framework. Future investigations could therefore focus on a more quantitative characterization of the thermo-induced nonlinear effects through explicit modelling of thermo-ionic transport, basin-of-attraction computations, and multistability investigations. Such approaches would provide deeper insight into the effective coexistence of distinct attractors and the long-term dynamical quantitative organization of thermo-controlled memristive electromechanical dynamics.

\section*{Funding Declaration}

 NGK appreciates financial support from the FAPESP--UNESCO-TWAS Project (Grant No. 2024/08346-8). 

 \section*{Acknowledgements}
 HAC thanks FAPESP grant 2021/14335-0 of the ICTP--SAIFR for partial support.\\

\section*{Declarations}
\subsection*{Conflict of interest / Competing interests}
The authors declare that they have no conflict of interest and no competing interests.

\subsection*{Ethics approval and consent to participate}
Not applicable.

\subsection*{Data availability}
 Data sets were not generated or analyzed during the current study. All results were obtained from numerical simulations using the original Python code developed by the authors.

\bibliography{sn-bibliography}
\section*{Appendix}
\begin{figure*}[!htb]
    \begin{subfigure}[b]{0.50\textwidth}
        \centering
        \includegraphics[width=7cm, height=5cm,trim=0cm 0cm 15.2cm 0cm,clip]{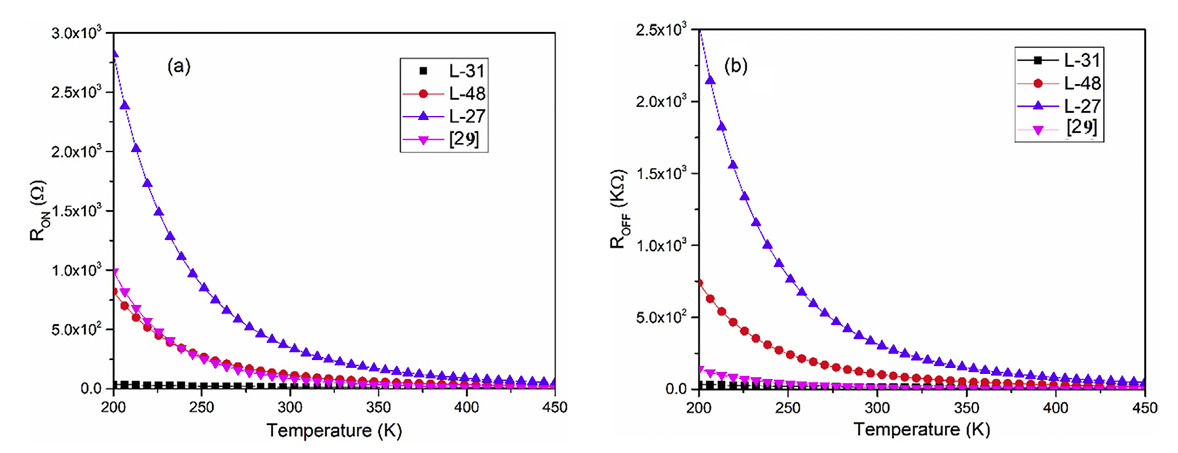}
        \subcaption{Influence of temperature on ON region’s resistance $R_{on}$.\cite{singh2018temperature}}
        \label{fig:graphe1}
    \end{subfigure}
    \hspace{1cm}
     \begin{subfigure}[b]{0.50\textwidth}
        \centering
       \includegraphics[width=7cm, height=5cm,trim=15.2cm 0cm 0cm 0cm,clip]{Variation_R.png}
        \subcaption{Influence of temperature on OFF region’s resistance $R_{off}$.\cite{singh2018temperature}}
        \label{fig:graphe1}
    \end{subfigure} 
    \begin{subfigure}[b]{0.50\textwidth}
        \centering{
        \includegraphics[width=7cm, height=5cm]{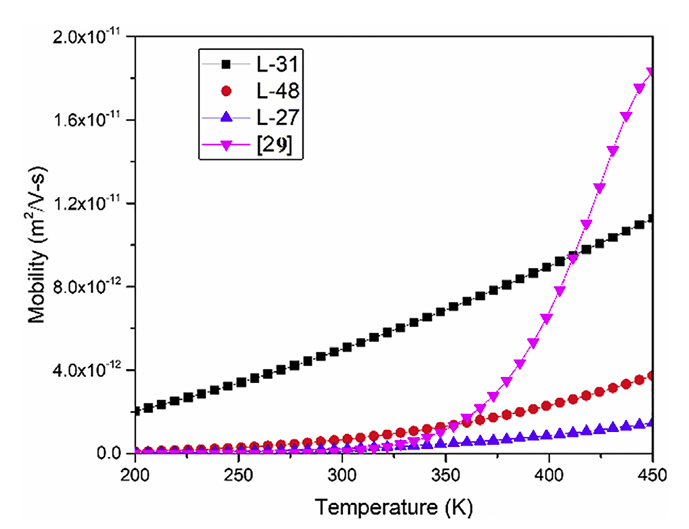}
        \subcaption{ Temperature dependent charge carrier mobility of memristor. \cite{singh2018temperature}}}
        \label{fig:graphe2}
    \end{subfigure}
    \caption{ Influence of temperature on,(a) ON region’s resistance (b) OFF region’s resistance  (c) Temperature dependent charge carrier mobility of memristor for three samples L-31,L-48,and L-27. Results are compared
with reported data \cite{garcia2016spice}.} 
    \label{fig14}
\end{figure*}

\end{document}